\documentclass[useAMS,usenatbib]{mn2e}

\usepackage{natbib}
\usepackage{graphicx}
\usepackage{multirow}
\usepackage{subfig}

\title[Nuclear $11.3\,\mu$m PAH emission in local
AGN]{Nuclear $11.3\,\mu$m PAH
  emission in local active galactic nuclei}
\author[Almudena Alonso-Herrero et al.]
{\parbox{\textwidth}{A.
  Alonso-Herrero,$^{1}\thanks{Augusto G. Linares Senior Research Fellow.}$\thanks{E-mail: aalonso@ifca.unican.es}
  C. Ramos Almeida,$^{2,3}$\thanks{Marie Curie Fellow.}
  P. Esquej,$^{4}$  P. F. Roche,$^5$  A.
    Hern\'an-Caballero,$^1$ S. F. H\"onig,$^6$\thanks{Marie Curie Fellow.} O. 
    Gonz\'alez-Mart\'{\i}n,$^{2,3}$  
I. Aretxaga,$^{7}$ 
R. E. Mason,$^{8}$   C. Packham,$^{9}$ N. A. Levenson,$^{10}$  
J. M. Rodr\'{\i}guez Espinosa,$^{2,3}$  R. Siebenmorgen,$^{11}$
M. Pereira-Santaella,$^{12,13}$ 
T. D\'{\i}az-Santos,$^{14}$    L. Colina,$^{13}$ C. Alvarez,$^{2,3}$ and C. M. Telesco$^{15}$}
\vspace{0.4cm} \\
$^{1}$Instituto de F\'{\i}sica de Cantabria, CSIC-UC, E-39005 Santander,
Spain\\
$^2$Instituto de Astrof\'{\i}sica de Canarias (IAC), E-38205 La Laguna,
Tenerife, Spain\\
$^3$Departamento de Astrof\'{\i}sica, Universidad de la Laguna (ULL),
E-38206 La Laguna, Tenerife, Spain\\ 
$^{4}$Departamento de Astrof\'{\i}sica, Universidad Complutense de
Madrid, E-28040 Madrid, Spain\\ 
$^{5}$Department of Physics, University of Oxford, Oxford OX1 3RH,
UK\\
$^6$Dark Cosmology Centre, Niels-Bohr-Institute, University of
Copenhagen, DK-2100 Copenhagen \O, Denmark\\
$^{7}$Instituto Nacional de Astrof\'{\i}sica, Optica y Electr\'onica (INAOE), 72000 Puebla, Mexico\\
$^{8}$Gemini Observatory, Northern Operations Center, Hilo,  HI 96720,
USA\\
$^{9}$Department of Physics and Astronomy, University of Texas at San
Antonio, San Antonio, TX 78249, USA\\
$^{10}$Gemini Observatory, Casilla 603, La Serena, Chile\\
$^{11}$European Southern Observatory, 
        D-85748 Garching b. M\"unchen, Germany\\
$^{12}$Istituto di Astrofisica e Planetologia Spaziali, INAF, I-00133 Roma, Italy\\
$^{13}$Centro de Astrobiolog\'{\i}a, CSIC-INTA, E-28850 Torrej\'on de
Ardoz, Madrid, Spain\\
$^{14}$Spitzer Science Center, Caltech, Pasadena, CA 91125, USA\\
$^{15}$Department of Astronomy, University of Florida, Gainesville, FL
32611, USA}

\begin{document}

\date{Accepted ---. Received ---; in original form ---}

\pagerange{\pageref{firstpage}--\pageref{lastpage}} \pubyear{2014}

\maketitle

\label{firstpage}

\begin{abstract}
We present Gran Telescopio CANARIAS CanariCam  
 $8.7\,\mu$m imaging and $7.5-13\,\mu$m spectroscopy of
six local systems known to host an active galactic
nucleus (AGN) and have nuclear
 star formation. Our main goal is to investigate whether
the molecules responsible for the $11.3\,\mu$m polyclyclic aromatic
hydrocarbon (PAH) 
feature are destroyed in the close vicinity of an AGN. We detect 
$11.3\,\mu$m PAH  feature emission in the nuclear regions 
of the galaxies  as well as extended PAH emission over
a few
hundred parsecs. The equivalent width (EW) of the feature shows a minimum  
at the nucleus but increases with increasing radial
distances, reaching typical 
star-forming values a few hundred parsecs away from the nucleus.   
The reduced nuclear EW are interpreted as due to increased dilution from the
AGN continuum rather than destruction of 
the PAH molecules.  
We conclude that at least those molecules responsible for
the $11.3\,\mu$m PAH feature survive in the nuclear environments as
close as 10\,pc from the AGN and for  
Seyfert-like AGN luminosities. 
We propose that material in the dusty tori, nuclear gas disks, and/or
host galaxies of AGN
is likely to provide the column densities necessary to protect the PAH
molecules from  the AGN radiation field.
\end{abstract}

\begin{keywords}
galaxies: individual: Mrk~1066, Mrk~1073, NGC~2273, Arp~299, NGC~6240,   
IRAS~17208$-$0014 -- galaxies: Seyfert -- infrared: galaxies --
galaxies: active
\end{keywords}

\section{Introduction}\label{sec:intro}

\begin{table*}
 \centering
 \begin{minipage}{175mm}
  \caption{Properties of the GTC/CanariCam sample.}\label{table:sample}
  \begin{tabular}{lcccccccccc}
 \hline
 Galaxy     & Optical  &$b/a$ &  z      & $D_{\rm L}$  & Scale     & IRAS $f_{12\mu{\rm
     m}}$ & $\log L_{\rm IR}$ & $L_{2-10{\rm keV}}$  & X-ray & Ref\\   
           & Class      &     &   &  (Mpc) & (pc/arcsec)   & (mJy)  & $(L_\odot)$& (erg
           s$^{-1}$) & \\ 
\hline
Mrk~1066 &        Sy2   & 0.6 &0.012025  & 47    & 224    & 460   & 10.91
& $7.8\times10^{42}$ & Corr & 1\\
Mrk~1073 &        Sy2$^*$ & 0.9  & 0.023343  & 95    & 442    & 440   & 11.39
& $1.5\times10^{41}$ & Obs & 2\\
NGC~2273 &        Sy2   &0.8  & 0.006138  & 26    & 124    & 440   & 10.15 
& $1.9\times10^{42}$ & Corr & 3 \\
Arp~299 & Sy2/L & 0.9 & 0.010300 & 44& 213 & 3970  & 11.83 & $1.9\times
10^{43}$$^{**}$ & Corr & 4 \\
NGC~6240 &   L   & 0.5 & 0.024480  & 103   & 475    & 590   & 11.85
& $3.0\times 10^{41}$$^{\S}$ (N) & Obs & 5\\
         &     &         &             &       &        &     &
& $1.0\times 10^{42}$$^{\S}$ (S) & Obs & 5\\
IRAS~17208$-$0014 & L & 0.8 & 0.042810  & 181   & 809    & 200   & 12.41
& $9.3\times10^{42}$ & Corr & 6 \\ 

\hline
\end{tabular}
Notes.---  Optical classifications are Sy=Seyfert and L=LINER.
Redshifts, axis ratios $b/a$, luminosity distances, and
projected 1\,arcsec \, scales in 
parsec are from NED.  
The {\it IRAS} $12\,\mu$m flux densities and IR ($8-1000\,\mu$m) luminosities
are from Sanders 
et al. (2003). The latter are corrected to the distances used in this
work. The X-ray  column indicates whether the hard X-ray luminosities
are  observed (Obs) or corrected (Corr) for intrinsic absorption, as
listed in their 
corresponding references: (1) Marinucci et al. (2012); (2) Guainazzi
et al. (2005); (3)  
Awaki et al. (2009); (4) Della Ceca et al. (2002), $^{**}$the luminosity is in
the $0.5-100$\,keV range; (5) Komossa et al. (2003), $^{\S}$the luminosities are in
the $0.1-10\,$keV range; (6)
Gonz\'alez-Mart\'{\i}n et al. (2009).\\
$^*$Detection of a broad component of Pa$\beta$
\citep{Veilleux1997}.
\end{minipage}
\end{table*}

The fueling of active galactic nuclei (AGN) requires that material
be driven inwards from the interstellar medium of the host galaxy
to physical scales of less than 1\,pc \citep[see the review by ][and
references therein]{Alexander2012}. Therefore, nuclear ($< 100\,$pc regions)
star formation appears to be an inevitable consequence of this
process. Indeed, numerical simulations predict that in AGN star
formation in the nuclear region  should be 
tightly correlated with the black hole accretion rate
\cite[e.g.][]{Kawakatu2008,Hopkins2010}. Moreover, 
nuclear star formation via stellar feedback
(stellar winds and supernovae) might help sustain the vertical
thickness of the optically and geometrically thick dusty torus of the
AGN unified model \citep[see 
e.g.][]{Wada2002,Vollmer2008}.

There is plenty of observational evidence for the presence of nuclear
star formation in AGN \citep{Davies2007,Esquej2014} and especially in
Seyfert 2s \citep{CidFernandes2001,GonzalezDelgado2001}. However, in
the local universe it
is not clear whether there is 
increased star formation activity in the hosts and nuclear
regions of type 2 AGN 
\citep[e.g.][]{Maiolino1995,Hicks2009} or not \citep{Clavel2000,Peeters2004},
and whether the 
nuclear/circumnuclear star formation level 
is a function  of the activity class \citep[e.g. radio galaxy, quasar,
infrared (IR) selected quasars, see][]{Shi2007} and/or the AGN luminosity
\citep{DiamondStanic2012, Esquej2014}. Furthermore, 
observational and theoretical arguments suggest the existence
of dynamical delays between the on-set of the
nuclear star formation and the fueling of the AGN
\citep{Davies2007, Wild2010, Hopkins2012}, perhaps indicating that these two
processes might not be coeval.

Classical indicators of the presence of on-going or recent star
formation (e.g. UV emission, 
H$\alpha$, Pa$\alpha$, [Ne\,{\sc ii}]$12.8\,\mu$m, modelling of
stellar populations) are difficult to use in the nuclear regions of
AGN, as they can be easily contaminated by bright AGN emission.
If we assume that  polycyclic aromatic 
hydrocarbon (PAH) features are excited by star formation activity in
AGN, they can be used as indicators of the star 
formation rate of galaxies based on their good correlation with the IR
luminosity \citep{Brandl2006,Smith2007}.
However,  \cite{Peeters2004} noted that PAH emission might be better
suited to trace B stars rather than O stars implying that PAH emission probes
star formation over time scales of a few tens of millions of years
rather than instantaneous star formation.  PAH
molecules are also predicted to be excited by AGN heating at distances
of approximately 100\,pc from the AGN in the case of optically thin
radiation \citep{Siebenmorgen2004}. However, for optically 
thick radiation the 
PAH features from star formation heating are much stronger than those
due to AGN 
heating \citep[see][for more details]{Siebenmorgen2004}. 

Emission from PAH features is observed in the nuclear/circumnuclear
regions of AGN 
\citep{Roche1991, Clavel2000, Laurent2000, Siebenmorgen2004, Deo2009,
  Wu2009, Sales2010, DiazSantos2010, Hoenig2010, Tommasin2010, 
  DiamondStanic2010,  DiamondStanic2012, GonzalezMartin2013,
  Esquej2014}. However, in general the PAH emission in AGN appears to
have a lower contrast compared to the continuum emission than in starburst 
galaxies. It has been argued theoretically
\citep{Voit1992,Siebenmorgen2004} and observationally
\citep{Aitken1985,Wu2009,Sales2010} that 
this might be due to destruction of the PAH molecules in the close vicinity
of the harsh radiation field of AGN. However, \cite{DiamondStanic2010}
showed that in local Seyferts it is only the 6.2, 7.7, and $8.6\,\mu$m
PAH features that are suppressed on kpc scales  with respect to the
$11.3\,\mu$m PAH feature \citep[see also][]{Smith2007}. 

Recently,
\cite{Esquej2014} found no 
evidence of destruction of the $11.3\,\mu$m PAH carriers in the
nuclear regions (typically the central 60\,pc) of  a sample of 29
local Seyferts as a function of the AGN luminosity. These authors
argued that it is likely that the PAH molecules are protected from the
AGN radiation field by
material in the dusty torus \citep[see also][]{Voit1992, Miles1994}.
\cite{Sales2013} came to a similar conclusion for the low
luminosity AGN in NGC~1808. \cite{RuschelDutra2014}, on the other
hand, interpreted the non-detection of the nuclear $11.3\,\mu$m PAH
feature in  two Seyfert nuclei as due to destruction of the PAH
molecules by the AGN radiation field.

The main goal of this work is to explore the behaviour of the $11.3\,\mu$m PAH 
feature and in particular whether the molecules responsible for this
feature are destroyed in the close vicinity of an
AGN.  To this end, we have obtained mid-infrared (mid-IR) imaging and
spectroscopy using CanariCam \citep{Telesco2003} on the 10.4\,m Gran Telescopio
CANARIAS (GTC) of a sample of local AGN with nuclear star formation.
The paper is organised as follows. Section~2 summarizes the
AGN and star 
formation properties of the galaxies. 
Section~3 presents the
GTC/CanariCam observations and data reduction as well as mid-IR
spectroscopy obtained with the IR spectrograph 
\citep[IRS, ][]{Houck2004} on board the 
{\it Spitzer Space Telescope} and Section~4 
the analysis of the data. Section~5 describes the results on the
extended continuum and $11.3\,\mu$m PAH feature. In Section~6
we discuss the EW of the $11.3\,\mu$m feature, the nuclear
spectra of Seyfert galaxies, and PAH survival in the close vicinity of AGN. Finally, we give
our conclusions in Section~7. Throughout this 
work we have assumed $H_0=73\,{\rm km \,s}^{-1}{\rm Mpc}^{-1}$,
$\Omega_\Lambda =0.73$, and $\Omega_{\rm M}=0.27$.

\section[]{The Sample}\label{sec:sample}
We are conducting a mid-IR imaging and spectroscopic survey of
approximately 100 local active galaxies using   GTC/CanariCam. The
sample includes high luminosity AGN (PG quasars), 
Seyfert galaxies, radio galaxies, and low 
luminosity AGN. Some of these are hosted by luminous and
ultraluminous IR galaxies (LIRGs and ULIRGs) with IR
luminosities of $L_{\rm IR}=10^{11}-10^{12}\,L_\odot$ and $L_{\rm
  IR}>10^{12}\,L_\odot$, respectively. The
sample covers almost six orders of magnitude in AGN 
luminosity. The GTC/CanariCam observations include 
imaging at $8.7\,\mu$m and $7.5-13\,\mu$m spectroscopy as well as
polarimetry for selected objects. 

Among the galaxies in our sample already observed 
with CanariCam (see Section~3.1),  we selected for this work six
systems for which there is evidence in the literature of AGN
activity as well as  nuclear and circumnuclear star
formation. Additionally, we chose galaxies showing extended
$11.3\,\mu$m PAH emission in the CanariCam spectra so we could
investigate the effects of the AGN on the behaviour of the feature in
the vicinity of the active nucleus. 
We list the main properties of the six systems chosen for this work
in Table~\ref{table:sample}. We use, when possible, the intrinsic hard
X-ray $2-10\,$keV  
luminosity as a proxy for the AGN luminosity. In what follows we
summarize evidence of nuclear and circumnuclear on-going
or recent star formation and the AGN 
properties for each of them. For Arp~299 (IC~694 + NGC~3690) we refer the
reader to Alonso-Herrero et al. (2013, and references therein).

\begin{table}
 \centering
 \begin{minipage}{80mm}
  \caption{Log of the GTC/CanariCam imaging
    observations.}\label{table:imaging_log} 

  \begin{tabular}{llccccccl}
 \hline
 Galaxy     & Date  & $t_{\rm on}$      & Standard  & IQ & \\   
            &       &   (s) &    & (arcsec)\\ 
\hline
Mrk~1066 &  2013.08.27 & 417 & HD18449 & 0.24 \\
Mrk~1073 &  2013.08.27 & 626 & HD19476 & 0.26 \\
NGC~2273 &  2013.09.24 & 626 & HD42633 & 0.26 \\
NGC~6240 &  2013.08.27 & 626 & HD151217 & 0.38 \\
IRAS~17208$-$0014 & 2013.06.07 & 1043 & HD157999 & 0.26\\
\hline
\end{tabular}
\end{minipage}
\end{table}

\begin{table*}
 \centering
 \begin{minipage}{90mm}
  \caption{Log of the GTC/CanariCam spectroscopic
    observations.}\label{table:spectroscopy_log} 

  \begin{tabular}{llccccccl}
 \hline
 Galaxy     & Date  & $t_{\rm on}$      & PA & Standard  & IQ  \\   
            &       &   (s) &   ($^{\circ}$) & & (arcsec) \\ 
\hline
Mrk~1066 &  2013.08.31 & 1061 & 315 & HD18449 & 0.28\\
Mrk~1073 &  2013.09.10 & 1238 & 75  & HD14146 & 0.34 \\
NGC~2273 &  2013.09.23 & 884  & 290 & HD42633 & 0.26 \\
         &  2013.09.22 & 354  & 290 & HD42633 & 0.32 \\
NGC~6240 &  2013.09.15 & 1238 & 16  & HD157999 & 0.40 & \\
IRAS~17208$-$0014 & 2013.09.09 & 1238 & 90 & HD157999 & 0.30\\
\hline
\end{tabular}
\end{minipage}
\end{table*}

 \begin{figure}

\hspace{0.5cm}
\resizebox{0.95\hsize}{!}{\rotatebox[]{-90}{\includegraphics{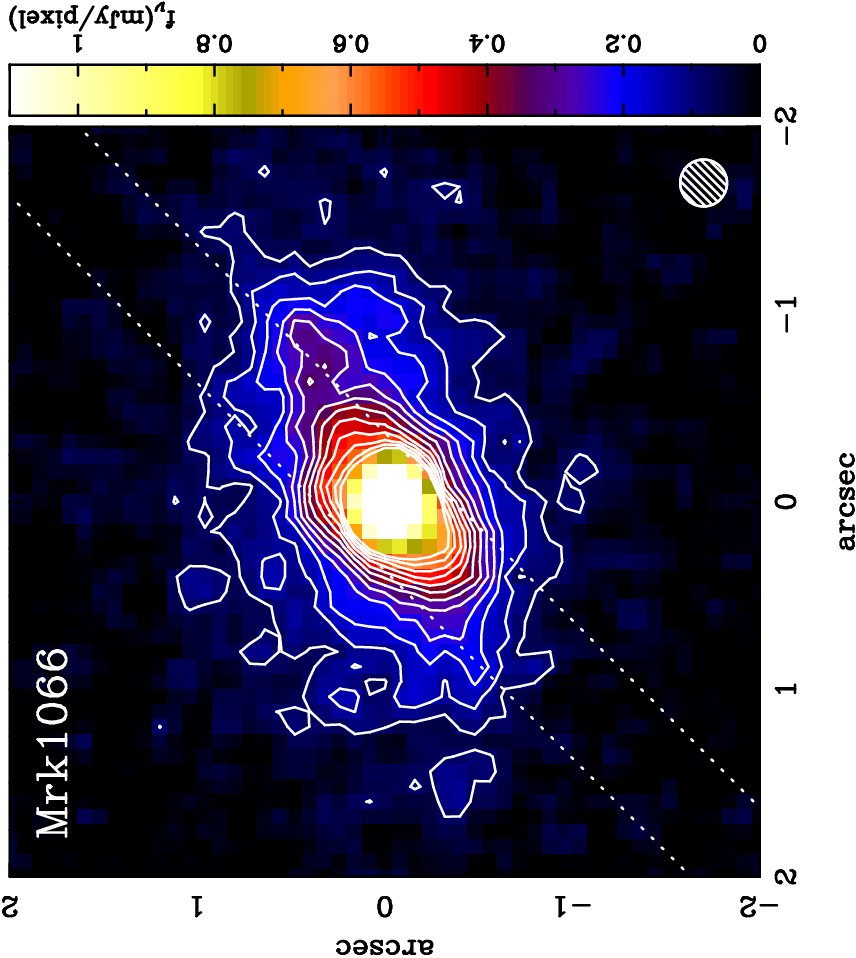}}}

\resizebox{1.\hsize}{!}{\rotatebox[]{-90}{\includegraphics{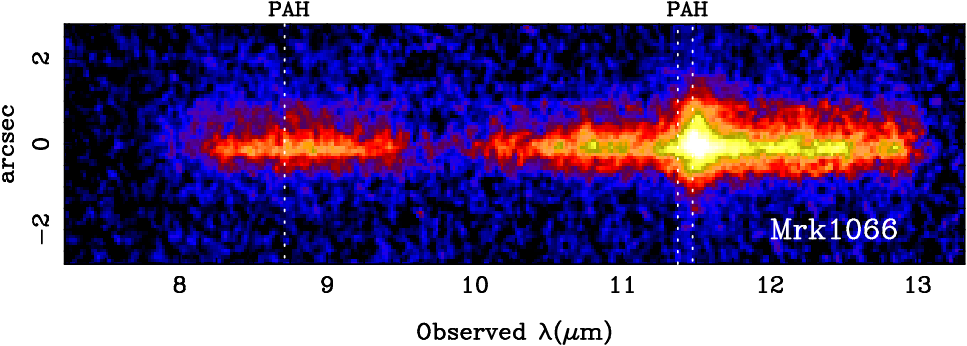}}}

\vspace{-2.5cm}

\resizebox{1.\hsize}{!}{\rotatebox[]{-90}{\includegraphics{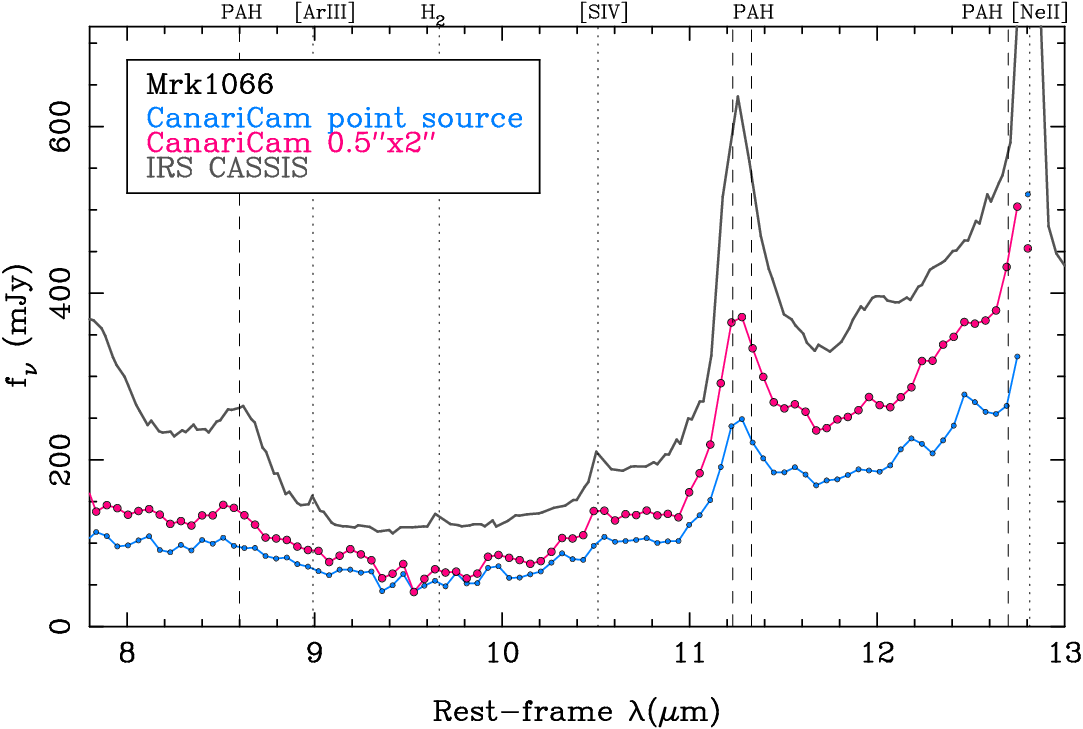}}}

\vspace{-1.5cm}
     \caption{Mrk~1066. The top panel is the GTC/CanariCam Si-2
       ($\lambda_{\rm c} = 8.7\,\mu$m) 
        image on a colour linear scale as plotted on the right hand
       side. Orientation is north 
       up, east to the left. The image was smoothed with a Gaussian
       function with  
       $\sigma =0.6\,$pixels. The contours are also in a linear scale
       with the lowest contour corresponding to the local background
       value plus 1.5 standard 
       deviations measured in the image before rotating and smoothing
       the images. The hatched circle represents the
       angular resolution of the image (FWHM) approximated as a
       Gaussian function. The dotted lines show the 
approximate width and orientation of the
       slit.  The middle panel shows the partially reduced
       GTC/CanariCam 2D spectrum in a square-root colour scale and smoothed
       with a Gaussian function.  On the y-axis 
       positive offsets indicate northwest and negative offsets
       southeast. The bottom panel compares the
     GTC/CanariCam nuclear and $0.52\,{\rm arcsec} \times 2\,{\rm arcsec}$ 
     spectra with the {\it Spitzer}/IRS SL spectrum extracted as a
     point source.  We
     smoothed the GTC/CanariCam spectra by averaging together three
     pixels in the spectral direction. 
}  
\label{fig:Mrk1066_images}
    \end{figure}

{\it Mrk~1066 (UGC~2456)} --- This nearly LIRG ($L_{\rm IR}\sim 8
  \times 10^{10}\,L_\odot$) is optically
classified as a Seyfert 2 galaxy with 
star formation activity in the nuclear region on scales of $\sim
1.5\,{\rm arcsec} = 340\,$pc
\citep{GonzalezDelgado2001,CidFernandes2001,RamosAlmeida2009b,Riffel2010}. In 
particular,  \cite{Riffel2010} used near-IR integral field unit (IFU)
observations to reveal the 
presence of two star forming knots at projected distances of 0.5\,arcsec \,
southeast and 1\,arcsec \, northwest from the nucleus, respectively.
However, at 
radial distances from the AGN of less than approximately 0.5\,arcsec 
the near-IR line ratios are dominated by AGN processes. This
galaxy also shows a nuclear molecular gas disk detected in the
$2.12\,\mu$m H$_2$ emission line. This gas disk  is 
believed to provide the material to feed the AGN \citep{Riffel2011}.
Mrk~1066 hosts a nearly 
Compton-thick Seyfert 2 nucleus \citep[X-ray column density of $N_{\rm H}=9\times
10^{23}\,{\rm cm}^{-2}$, ][]{Guainazzi2005} with a large observed
equivalent width (EW) of the
Fe K$\alpha$ line at 6.4\,keV. 
\cite{Marinucci2012} estimated an absorption corrected intrinsic
$2-10\,$keV luminosity of $7.8\times 10^{42}\,{\rm erg \, s}^{-1}$.  

 \begin{figure}

\hspace{0.5cm}
\resizebox{0.95\hsize}{!}{\rotatebox[]{-90}{\includegraphics{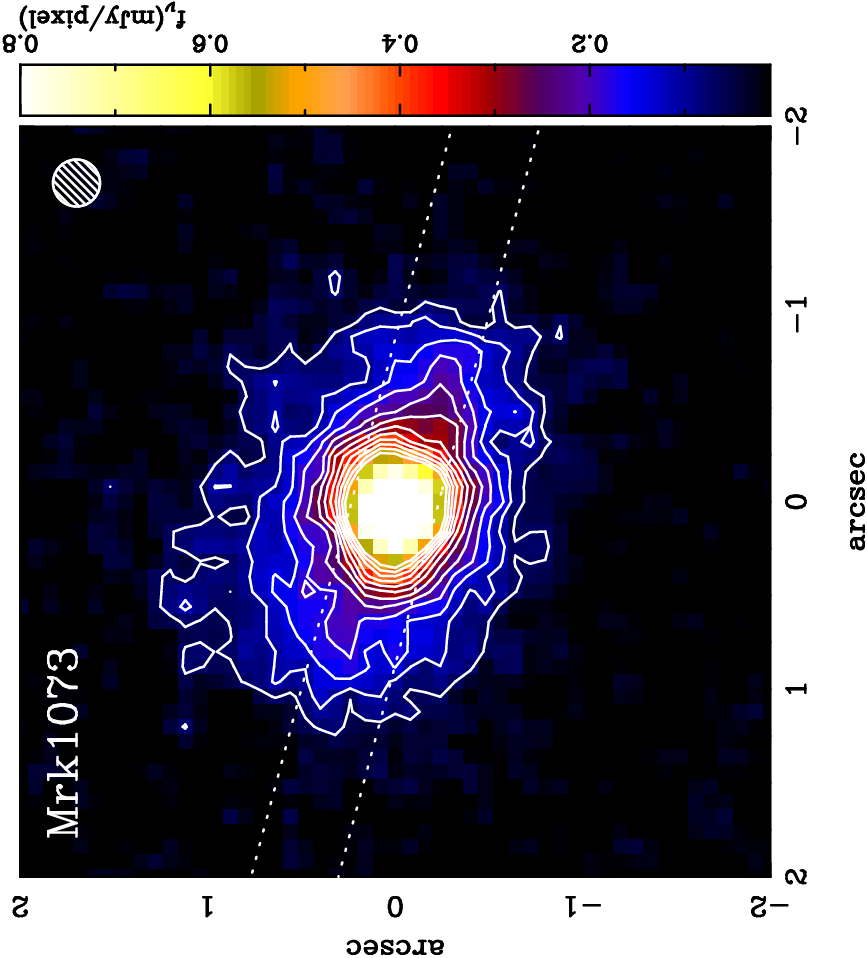}}}

\resizebox{1.\hsize}{!}{\rotatebox[]{-90}{\includegraphics{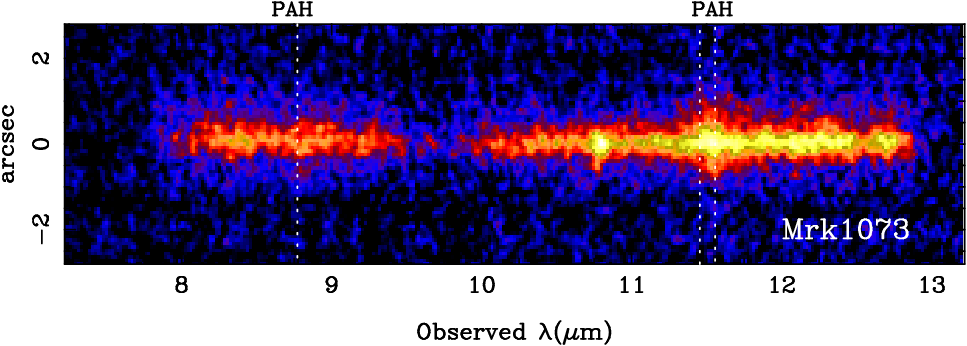}}}

\vspace{-2.5cm}

\resizebox{1.\hsize}{!}{\rotatebox[]{-90}{\includegraphics{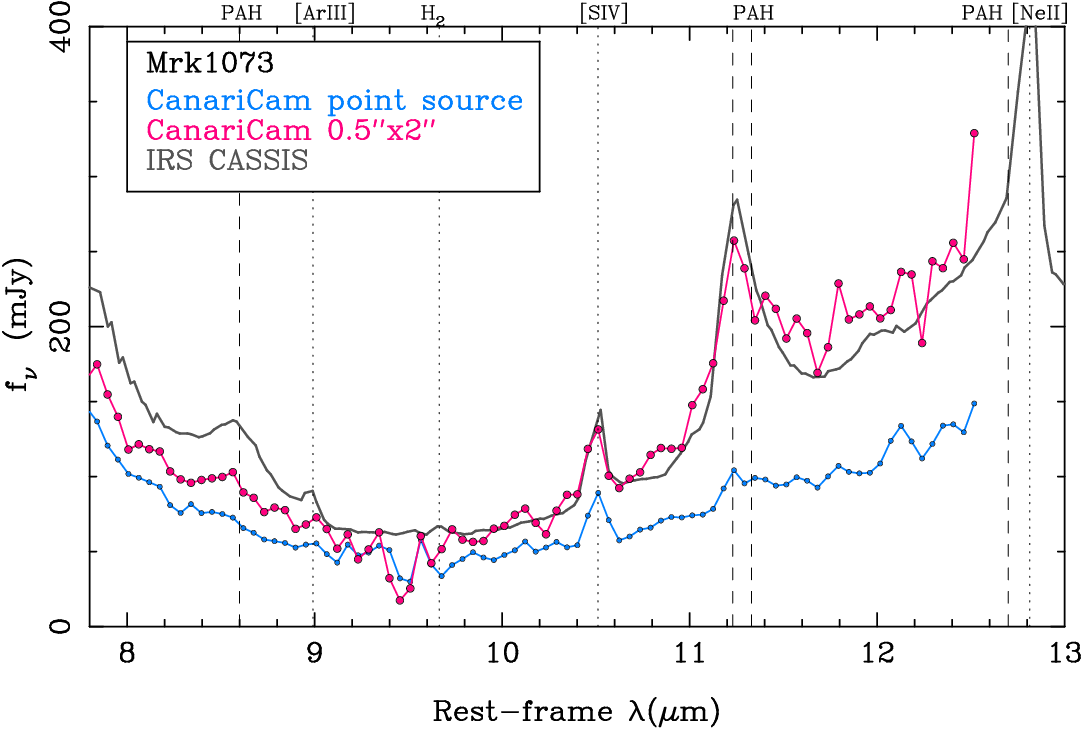}}}

\vspace{-1.5cm}
   \caption{Mrk~1073. Symbols as in
     Fig.~\ref{fig:Mrk1066_images}. The top panel is the GTC/CanariCam
     $8.7\,\mu$m 
     image smoothed with a Gaussian function with 
     $\sigma =0.6\,$pixels.  The contours are in a linear scale
       with the lowest contour corresponding to the local background
       value plus 2 standard 
       deviations measured in the image before rotating and smoothing
       the images. The middle panel is the partially
     reduced GTC/CanariCam 2D 
     spectrum. On the y-axis positive
     offsets indicate  northeast and negative offsets
      southwest. 
The bottom panel compares the 1D GTC/CanariCam spectra 
(smoothed as in
     Fig.~\ref{fig:Mrk1066_images}, bottom panel) and the
  {\it Spitzer}/IRS 
spectrum. The GTC/CanariCam $0.52\,{\rm arcsec} \times
2\,{\rm arcsec}$ 
     spectrum of Mrk~1073 was scaled down by a factor of 1.8 to match
     the IRS one.  }    
   \label{fig:Mrk1073_images}
\end{figure}

 \begin{figure}

\hspace{0.5cm}
\resizebox{0.95\hsize}{!}{\rotatebox[]{-90}{\includegraphics{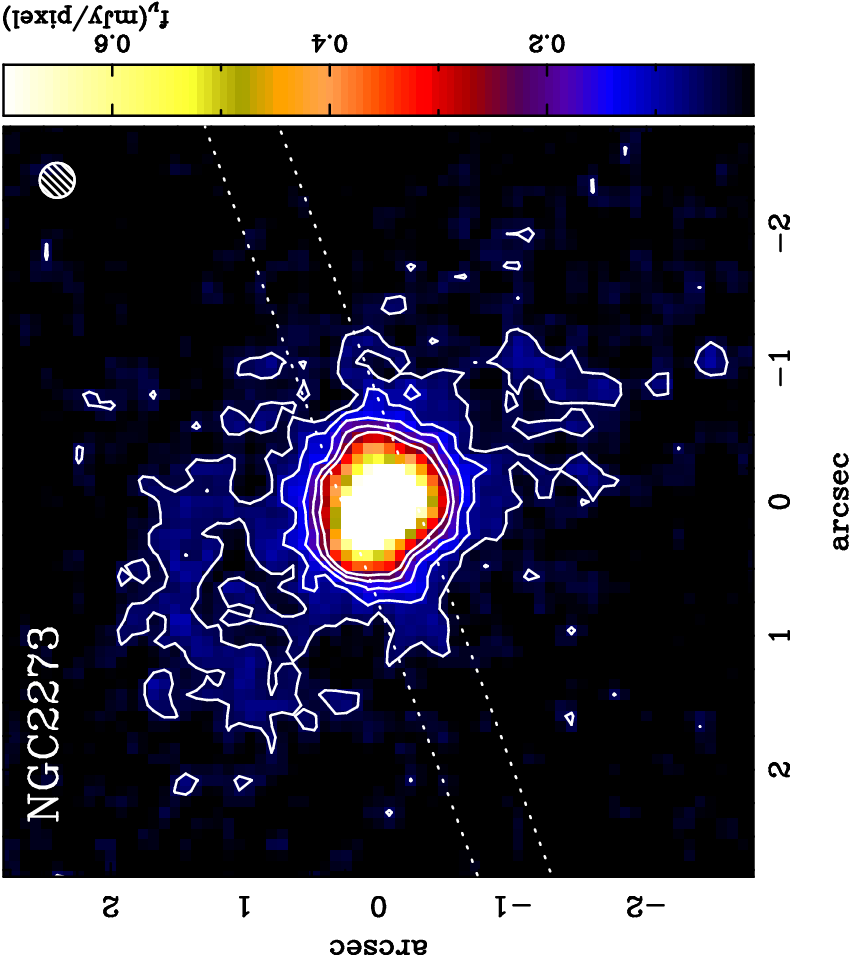}}}

\resizebox{1.\hsize}{!}{\rotatebox[]{-90}{\includegraphics{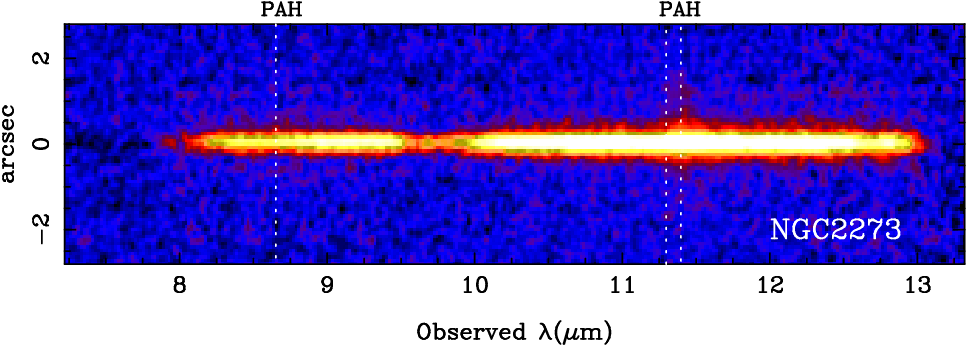}}}

\vspace{-2.5cm}

\resizebox{1.\hsize}{!}{\rotatebox[]{-90}{\includegraphics{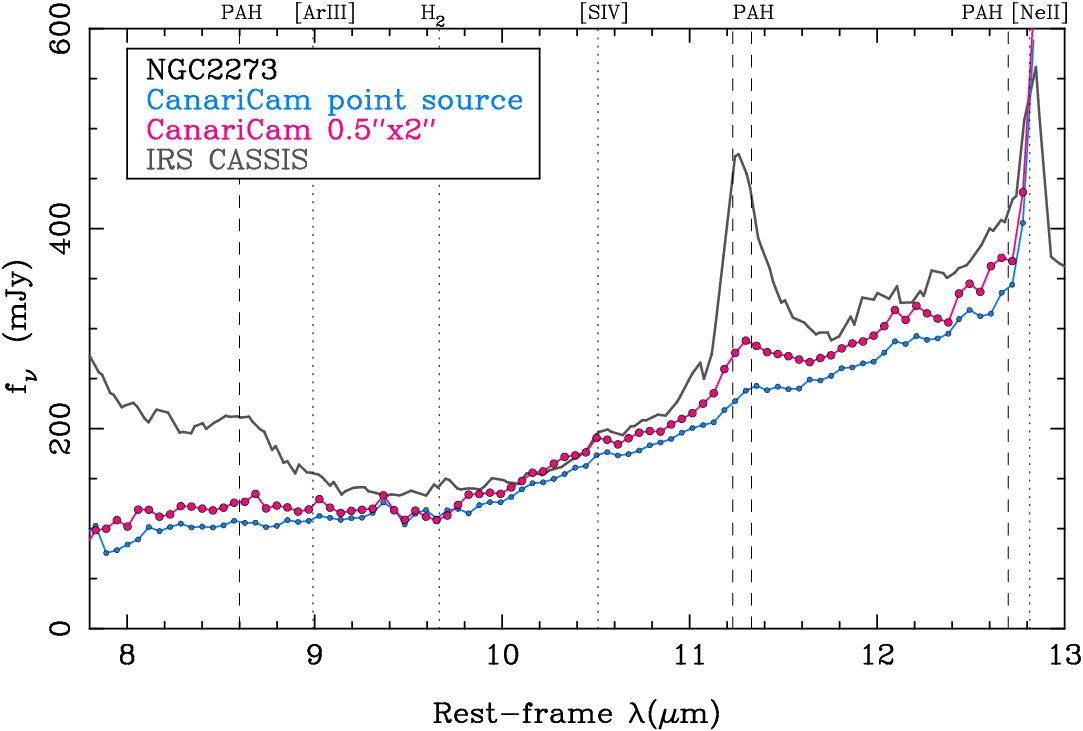}}}

\vspace{-1.5cm}
   \caption{NGC~2273. Symbols as in
     Fig.~\ref{fig:Mrk1066_images}. The top panel is the GTC/CanariCam
     $8.7\,\mu$m 
     image smoothed with a Gaussian function with 
     $\sigma =0.7\,$pixels.  The contours are in a linear scale
       with the lowest contour corresponding to the local background
       value plus 1 standard 
       deviations measured in the image before rotating and smoothing
       the images.  The middle panel is the partially 
reduced GTC/CanariCam 2D spectrum. On the
     y-axis positive offsets indicate northwest and negative offsets
       southeast. The bottom panel compares the 1D GTC/CanariCam 
spectra (smoothed as in
     Fig.~\ref{fig:Mrk1066_images}, bottom panel)  and the 
         {\it Spitzer}/IRS 
spectrum.  }    
   \label{fig:NGC2273_images}
\end{figure}

{\it Mrk~1073 (UGC~2608)} --- This LIRG is located in the 
Perseus cluster and it is optically  classified as a Seyfert 2,
although it shows a broad component detected in the near-IR Pa$\beta$ line
\citep{Veilleux1997}. Mrk~1073 
was also included in a radio selected sample of starburst galaxies
\citep{Smith1996}. Using optical spectroscopy 
\cite{GonzalezDelgado2001} and \cite{CidFernandes2001} demonstrated that this
galaxy experienced  starburst  activity in the central $\sim
1.5\,{\rm arcsec}= 660\,$pc.  
 Mrk~1073 is a Compton-thick \citep[$N_{\rm H}>1.6\times
10^{24}\,{\rm cm}^{-2}$, ][]{Guainazzi2005} Seyfert 2 with a large
observed EW of the 
Fe K$\alpha$  line. We followed \cite{Marinucci2012} to correct the
observed $2-10\,$keV luminosity (see Table~\ref{table:sample}) and 
estimated an intrinsic $2-10\,$keV luminosity of $\sim
1\times10^{43}\,{\rm erg \, s}^{-1}$.

{\it NGC~2273} --- This Seyfert 2 galaxy hosts a circumnuclear ring of
star formation clearly detected in H$\alpha$+[N\,{\sc ii}] line emission with an approximate
diameter of $\sim 4\,$arcsec (i.e. 500\,pc), while the nuclear optical line
ratios are more compatible 
with AGN emission \citep{Ferruit2000}.  From optical Gemini/GMOS IFU
spectroscopy,
there is also  evidence of the presence of a nuclear ring of star
formation about $2-3\,$arcsec in diameter, based on the presence of a nuclear
velocity dispersion drop \citep{Barbosa2006}. This galaxy shows
nuclear dense gas emission but 
based on the derived Toomre parameter, \cite{Sani2012} concluded that star
formation is unlikely to be taking place in the nuclear region. Using
{\it Suzaku} observations, \cite{Awaki2009} confirmed the Compton-thick
nature of this Seyfert galaxy and derived an absorption corrected $2-10\,$keV
luminosity $1.9\times 10^{42}\,{\rm erg \,s}^{-1}$ (for our assumed distance).

{\it NGC~6240.} ---  This interacting nearly ULIRG hosts two X-ray
identified AGN at a projected separation of approximately $1.5\,{\rm arcsec}
= 715\,$pc \citep{Komossa2003}.  From a detailed modelling of the
adaptive optics (AO) near-IR VLT/SINFONI IFU observations, \cite{Engel2010} concluded
that there is recent merger-induced star formation activity in both
nuclei. However, the young starbursts do not dominate the near-IR  luminosity
of the system. Using Keck mid-IR spectroscopy \cite{Egami2006} showed
that the $11.3\,\mu$m PAH emission in the nuclear regions is extended over
$\sim 3\,{\rm arcsec} =1.4\,$kpc. The {\it Chandra}
X-ray observations of the nuclei revealed the presence of Fe K$\alpha$
lines with high EW indicating the Compton-thick nature of the 
nuclei. The observed $0.1-10\,$keV luminosities of the
southern and northern nuclei, are approximately $10^{42}\,{\rm erg \,
  s}^{-1}$ and $ 3 \times 10^{41}\,{\rm erg \,
  s}^{-1}$ \citep{Komossa2003}. The absorption corrected $2-10\,$keV X-ray
luminosities are therefore expected to be a few times $10^{43}\,{\rm erg \,
  s}^{-1}$. By modelling the IR emission of the southern
nucleus using clumpy torus models, \cite{Mori2014}
estimated that the AGNs contribute to $30-50\%$ of the total luminosity
of the system.

{\it IRAS~17208$-$0014} --- This is a late-stage merger ULIRG with a
single nucleus. \cite{Arribas2003} compared optical IFU observations
with {\it Hubble Space Telescope (HST)} near-IR data and concluded
that the optical peak does not 
coincide with the true near-IR nucleus. They also reclassified the
nuclear activity to LINER. There is evidence of nuclear ($\sim
250\,$pc) and extended star formation
from Keck mid-IR imaging \citep{Soifer2000}. 
\cite{PiquerasLopez2012} obtained VLT/SINFONI
near-IR IFU observations of IRAS~17208$-$0014 and found extended Pa$\alpha$
emission over $4\,{\rm arcsec} = 3.2\,$kpc with 
an off-nuclear region with a
large EW of Pa$\alpha$ and 
located approximately 1\,kpc to the southeast. 
The hard X-ray emission of this
galaxy is explained by the presence of a Compton-thick AGN with an
intrinsic $2-10\,$keV luminosity of $9 \times 10^{42}\,{\rm erg \,
  s}^{-1}$ \citep{GonzalezMartin2009}. 

 \begin{figure}

\hspace{0.5cm}
\resizebox{0.95\hsize}{!}{\rotatebox[]{-90}{\includegraphics{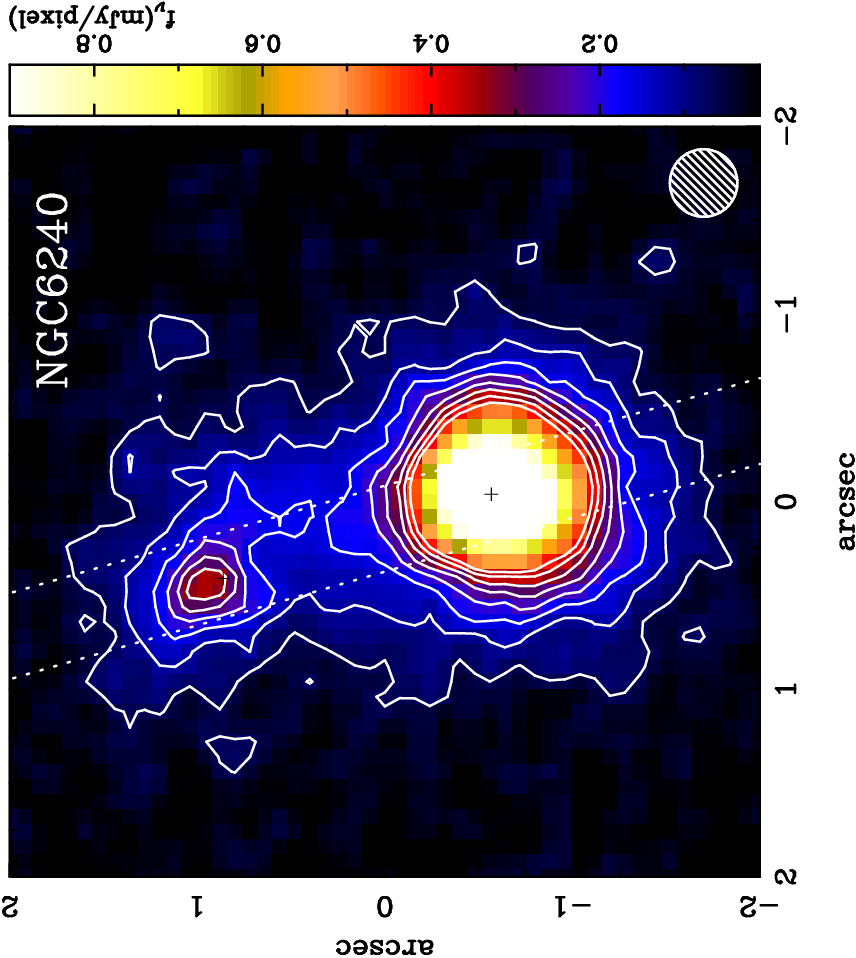}}}
\resizebox{1.\hsize}{!}{\rotatebox[]{-90}{\includegraphics{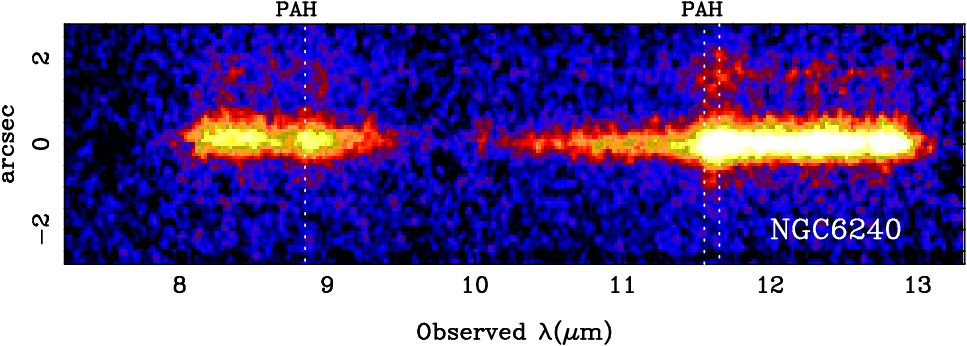}}}
\vspace{-2.5cm}

\resizebox{1.\hsize}{!}{\rotatebox[]{-90}{\includegraphics{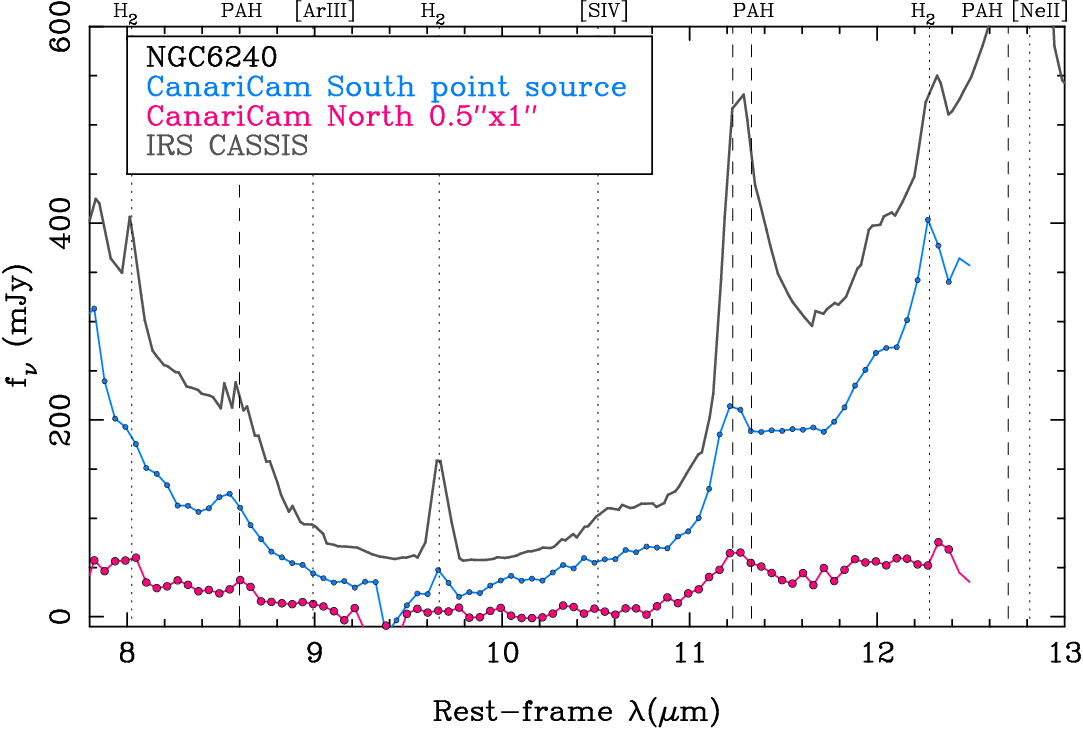}}}

\vspace{-1.5cm}
   \caption{NGC~6240. Symbols as in
     Fig.~\ref{fig:Mrk1066_images}. The top panel is the GTC/CanariCam
     $8.7\,\mu$m 
     image smoothed with a Gaussian function with 
     $\sigma =0.7\,$pixels.  The contours are in a linear scale
       with the lowest contour corresponding to the local background
       value plus 1.5 standard 
       deviations measured in the image before rotating and smoothing
       the images. The crosses indicate the
     positions of the radio sources \citep{Beswick2001} 
     after we made the radio southern source
     coincident with the bright mid-IR southern nucleus.  
    The middle panel is the partially reduced GTC/CanariCam 2D
      spectrum. On 
     the y-axis positive offsets indicate
     northeast and negative  offsets southwest. The northern nucleus is
     located at approximately $+1.5\,$arcsec from the southern
     nucleus. Bottom panel are the 1D GTC/CanariCam spectra (smoothed as in
     Fig.~\ref{fig:Mrk1066_images}, bottom panel) through 
     extraction apertures centered on the southern and northern nuclei
     compared to the {\it Spitzer}/IRS SL spectrum, which encompasses
     both nuclei.}   
   \label{fig:NGC6240_images}
\end{figure}

 \begin{figure}

\hspace{0.5cm}
\resizebox{0.95\hsize}{!}{\rotatebox[]{-90}{\includegraphics{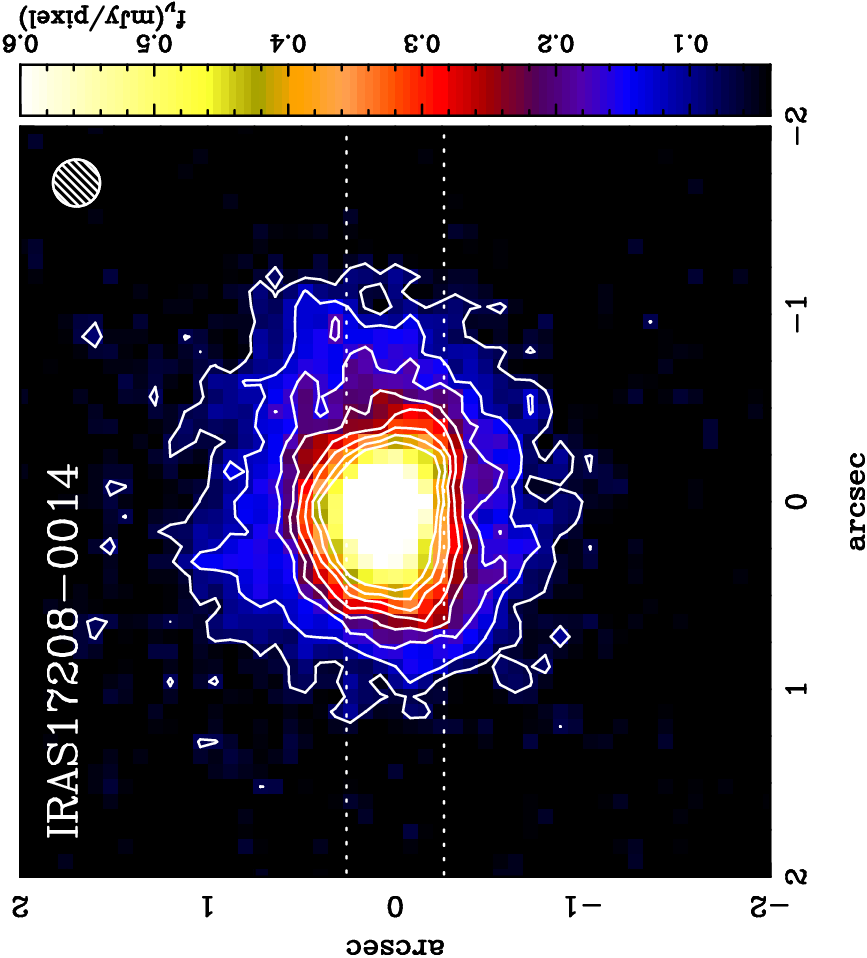}}}

\resizebox{1.\hsize}{!}{\rotatebox[]{-90}{\includegraphics{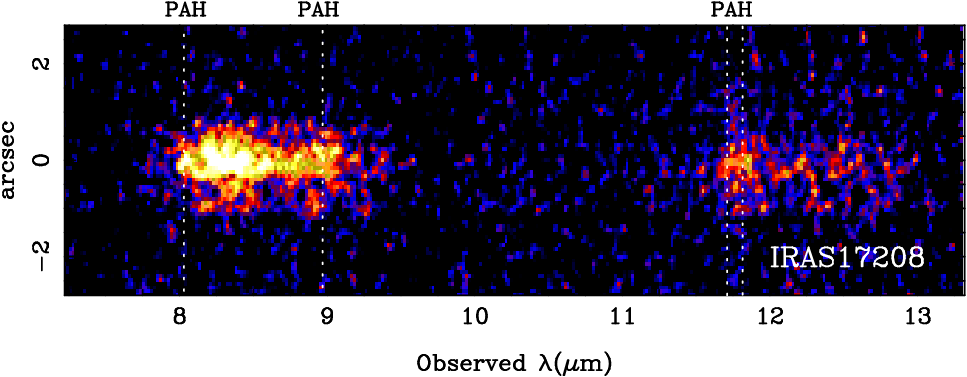}}}

\vspace{-2.5cm}

\resizebox{1.\hsize}{!}{\rotatebox[]{-90}{\includegraphics{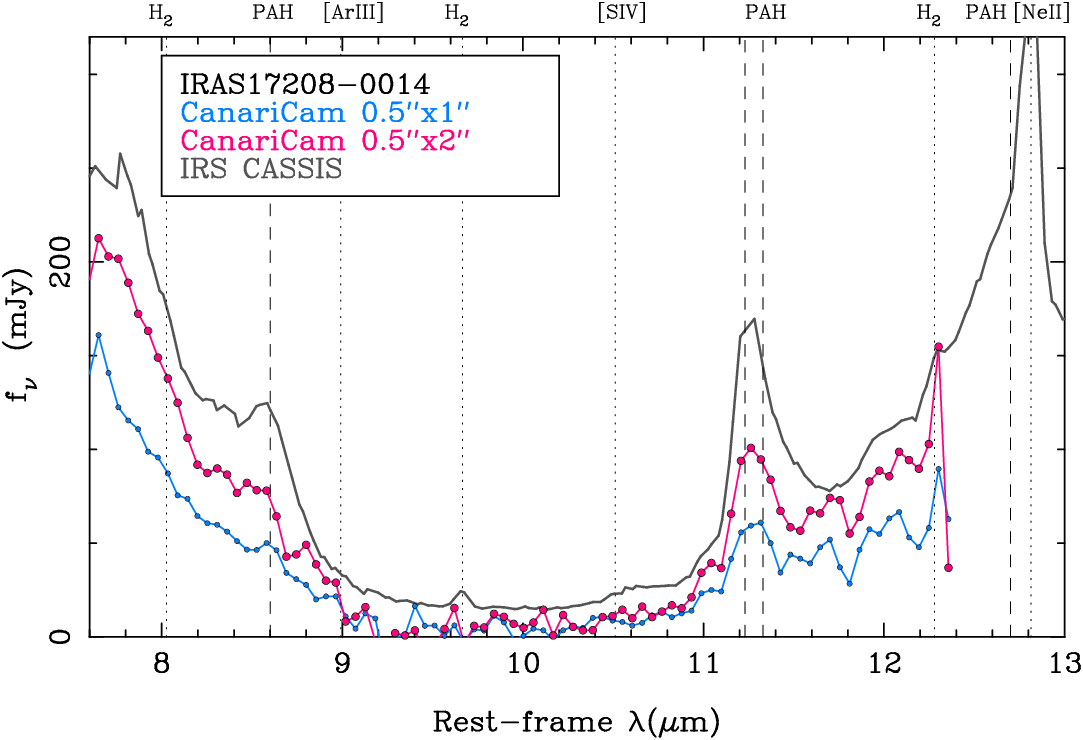}}}

\vspace{-1.5cm}
   \caption{IRAS~17208$-$0014. Symbols as in
     Fig.~\ref{fig:Mrk1066_images}. 
The top panel is the GTC/CanariCam $8.7\,\mu$m
     image smoothed with a Gaussian function with 
     $\sigma =0.6\,$pixels.  The contours are in a linear scale
       with the lowest contour corresponding to the local background
       value plus two standard 
       deviations measured in the image before rotating and smoothing
       the images. The middle panel is the partially
     reduced GTC/CanariCam 2D 
     spectrum. On the y-axis positive
     offsets indicate east and negative offsets 
       west. The bottom panel compares the 1D GTC/CanariCam 
spectra (smoothed as in
     Fig.~\ref{fig:Mrk1066_images}, bottom panel) 
and the {\it
           Spitzer}/IRS spectrum.  }    
   \label{fig:IRAS17208_images}
\end{figure}

Summarizing, the six systems selected for this work, even those
  optically classified as LINER (see Table~\ref{table:sample}), show AGN
  luminosities that are typical of the Seyfert galaxies in our
  CanariCam sample. In terms of their IR luminosities, all except
  NGC~2273, are nearly LIRGs, LIRGs or ULIRGs. This is well understood
  because we chose them 
  to have strong nuclear and circumnuclear star formation activity as
  most (U)LIRGs, and  in this sense they may not be 
  representative of the entire AGN sample. We will present a 
full study of the nuclear star formation activity in our sample of AGN
using the $11.3\,\mu$m PAH feature once the CanariCam observations are
finalised. 
\clearpage

\section[]{Observations and Data Reduction}\label{sec:observations}

\subsection{GTC/CanariCam observations}

The new CanariCam observations in this work 
were taken as  part of an ESO/GTC large programme (182.B-2005),
which totals 180
hours of observing time. At the time of writing approximately 75\% of
the observations have already been taken. We have an additional $\sim 100$
hours for this  project through the 
CanariCam guaranteed time programme. 

We obtained imaging
observations of the galaxies using 
the Si-2 filter ($\lambda_{\rm 
  c}=8.7\,\mu$m and width $\Delta \lambda_{\rm cut} =1.1\,\mu$m at 50\%
cut-on/off). The observations were taken in queue mode under
photometric conditions using the  standard mid-IR chop-nod
technique. The chop and nod throws were 15\,arcsec, whereas the chop
and nod angles were chosen for each target to avoid extended galaxy
emission in the sky image. We 
also observed standard stars immediately before or  
after the galaxy observations to perform the photometric calibration, 
measure the image quality (IQ) of the observations, and perform the point
spread function (PSF) subtraction. 

The plate scale of the
CanariCam $320\times240$  Si:As detector is 
0.0798\,arcsec/pixel, which provides a field of view in imaging mode of
$\sim 26\,{\rm arcsec} \times 19\,{\rm arcsec}$. We measured the IQ of
the data by 
fitting the observations of the 
standard stars with a Gaussian function and obtained values of
$0.2-0.4\,$arcsec \, (full width half maximum, FWHM). In
Table~\ref{table:imaging_log} we summarize details of the imaging
observations, including the date of the observations, the on-source
integration time 
$t_{\rm on}$, the standard star used, and the IQ of the observations.

We also obtained long-slit spectroscopy  of the nuclear regions of
the galaxies using the low spectral resolution
$10\,\mu$m grating, which covers the $N$-band $\sim 7.5-13\,\mu$m with
a nominal spectral resolution of $R=\lambda / \Delta \lambda \sim
175$. We used a  
0.52\,arcsec  wide slit for all the galaxies with the position angles
(PA) of the slits, measured from the north to the east, 
listed in Table~\ref{table:spectroscopy_log}.  The observing sequence
was first to  take an acquisition image of the galaxy 
with the Si-2 filter, then place the slit, and finally integrate for
the on-source times given in Table~\ref{table:spectroscopy_log}. The
chop-nod parameters were as for the imaging observations. We
also observed standard stars using the same observing sequence to
provide the photometric calibration, the telluric correction, and the slit 
loss correction. We also used the acquisition images of the standard stars to 
obtain the IQ of the observations (see Table~\ref{table:spectroscopy_log}).

\begin{table*}
 \centering
 
 \begin{minipage}{140mm}
  \caption{{\it Spitzer}/IRS spectroscopy.}\label{table:spectroscopy_Spitzer} 

  \begin{tabular}{llccccccl}
 \hline
 Galaxy     & Program ID  & PI      & Cycle & Mode \\   
\hline
Mrk~1066          & 30572  & V. Gorjian  & 6 & Staring\\
Mrk~1073          & 30323  & L. Armus    & 6 & Staring\\
NGC~2273          & 86     & M. Werner   & 1 & Staring\\
Arp~299           & 21     & J. R. Houck & 1 & Spectral Mapping\\
NGC~6240          & 105    & J. R. Houck & 1 & Staring\\
IRAS~17208$-$0014 & 105    & J. R. Houck & 1 & Staring\\
\hline
\end{tabular}
\end{minipage}
\end{table*}

We reduced the data using the CanariCam pipeline {\sc redcan},  which
is thoroughly described by  
\cite{GonzalezMartin2013}. Briefly, the reduction process of 
the imaging data includes sky subtraction, stacking of the individual
images, and rejection of bad images. The flux calibration of the
galaxy images is done using the 
observations of the standard stars. We show the fully reduced GTC/CanariCam 
$8.7\,\mu$m images  of the galaxies of our sample in the top panels
of Figs~\ref{fig:Mrk1066_images} through
\ref{fig:IRAS17208_images}.

\begin{table*}
 \centering
 \begin{minipage}{140mm}
  \caption{GTC/CanariCam aperture photometry and unresolved nuclear fluxes at
    $8.7\,\mu$m.}\label{table:aperture_photometry}  

  \begin{tabular}{llccccccc}
 \hline
 Galaxy     & \multicolumn{2}{c}{FWHM} &  \multicolumn{5}{c}{Fluxes}  \\ 
            & \multicolumn{2}{c}{Nucleus} & 0.48\,arcsec & 0.96\,arcsec
            & 1.92\,arcsec & 4\,arcsec  & Unresolved\\    
            &  (arcsec)     &   (pc) &   (mJy) & (mJy) & (mJy) &
            (mJy) & (mJy) \\ 
\hline
Mrk~1066    & 0.30 & 67  & $41\pm 1$ & $79\pm 3$ & $135\pm10$ &
$193\pm 30$ & 75\\
Mrk~1073    & 0.33 & 146 & $30\pm 1$ & $60\pm 1$ & $109\pm5$ & $160\pm
20$ & 57\\
NGC~2273    & 0.30 & 37  & $53\pm 1$ & $94\pm 2$ & $132\pm8$ & $189\pm
30$ &113   \\
NGC~6240N   & 0.45 & 214 & $8\pm 1$ & $20\pm4$ & --- & ---  & --\\
NGC~6240S   & 0.38 & 181 & $45\pm 1$ & $98\pm4$ & --- & $231\pm
19^{*}$ & 165 \\
IRAS~17208$-$0014 & 0.52 & 421 & $23\pm 1$ & $55\pm 1$ & $105\pm3$ &
$151\pm 8$ & --\\ 
\hline
\end{tabular}
Notes.--- The listed apertures are diameters. The quoted
errors in the photometry are only due to uncertainties in the background 
estimate. $^{*}$The 4\,arcsec-diameter photometry for NGC~6240 is for an
aperture centered between the two nuclei.
\end{minipage}
\end{table*}

For the spectroscopy the first three steps of the data
reduction are the same as for the imaging. 
Additionally {\sc redcan} performs 
 the two-dimensional (2D) wavelength calibration of the galaxy and standard star
spectra using sky lines. Finally, the trace determination is done using the
standard star data. 
Figs~\ref{fig:Mrk1066_images} through \ref{fig:IRAS17208_images}
(middle panels)
show 2D spectral images after the wavelength
calibration but before correction for atmospheric transmission. From
these figures we can clearly observe  
nuclear and extended
emission of the $11.3\,\mu$m PAH feature in all five galaxies. 
The last steps of the spectroscopic data reduction are the spectral
extraction either as point sources or extended sources (see
Section~\ref{sec:spectra}), and finally, the correction for slit losses in the 
case of point source extractions.

\subsection{Spitzer/IRS spectroscopy}
All the galaxies in our sample were observed with {\it Spitzer}/IRS using
the short-low  
(SL) spectral resolution ($R \sim 60-120$)
module covering the spectral range $\sim 5-15\,\mu$m. 
  Table~\ref{table:spectroscopy_Spitzer} summarizes  information about the
  observations. All the observations were taken in staring mode, except
  for those of Arp~299 which were taken using the IRS spectral mapping
  mode capability. \cite{AlonsoHerrero2009} and \cite{AlonsoHerrero2013}
  provide all the details  
on the data reduction and extraction of the nuclear spectra of Arp~299
as a point sources, respectively. 
For the rest of the galaxies observed in staring mode,  
we downloaded the fully-calibrated  spectra from the Cornell Atlas of
Spitzer/IRS Sources \citep[CASSIS v4; ][]{Lebouteiller2011}.  
CASSIS  provides spectra with
optimal extraction regions to ensure the best signal-to-noise (S/N)
ratio. For the galaxies in this work the optimal CASSIS extraction was
  equivalent to 
a point source extraction. The SL slit width of these observations is
3.7\,arcsec.

\section{Analysis}
\subsection{Imaging: source size and aperture photometry}\label{sec:phot}
We measured the size of the nuclear regions before rotating and
smoothing the GTC/CanariCam images using a Gaussian function. 
The measured sizes (FWHM) of the mid-IR emission in arcseconds and
pc for the nuclear 
regions of the galaxies are listed in Table~\ref{table:aperture_photometry}.  
The nuclear sizes ($0.3-0.4\,$arcsec) at $8.7\,\mu$m of
Mrk~1066, Mrk~1073, NGC~2273, and NGC~6240S are consistent with the
presence of an unresolved source, although there is also extended
emission. The
nuclear region of IRAS~17208$-$0014 and NGC~6240N
appear clearly resolved at $8.7\,\mu$m in good agreement with previous
observations for instance at near-IR wavelengths \citep[see
e.g.,][respectively]{PiquerasLopez2012,Engel2010}.

We performed aperture photometry on the images using {\sc
  iraf}\footnote{IRAF is distributed by the National Optical Astronomy
  Observatory, which is operated by the Association of Universities
  for Research in Astronomy (AURA) under cooperative agreement with
  the National Science Foundation.} routines. The  
nuclear $8.7\,\mu$m flux densities (without the correction for point source
emission) measured through different apertures are given in
Table~\ref{table:aperture_photometry}.   We also list in this table
the uncertainties due to the background subtraction, which increase
with the size of the aperture used. Additionally the
typical errors of the photometric calibration in the $N$-band window
are estimated to be approximately $10-15\%$. 

For those galaxies with nuclear FWHMs similar to those of their
corresponding standard stars 
we also estimated the unresolved nuclear fluxes. To do so, we used the
fluxes measured through the smallest aperture and estimated the
aperture correction for the total flux using the standard star
observations. We note that the unresolved nuclear fluxes estimated
using this method (see Table~\ref{table:aperture_photometry}) are likely
to be slightly overestimated as there is always a 
small fraction of resolved emission from the galaxy even in the
smallest aperture. This is the case for Mrk~1066, where \cite{RamosAlmeida2014}
estimated an unresolved nuclear flux by PSF scaling of $63\pm 9\,$mJy.

\subsection{Extracting spectra}\label{sec:spectra}
For all the galaxies except NGC~6240N and IRAS~17208$-$0014, which
appear resolved in the GTC/CanariCam images at $8.7\,\mu$m (see
previous section and Table~\ref{table:aperture_photometry}), we
extracted the nuclear spectra as point sources. In this case {\sc
  redcan} uses an extraction aperture that increases with wavelength
to take care of the decreasing angular resolution. It also 
performs a correction to account for slit losses. We checked the flux
calibration of the nuclear spectra against the photometry done on the
GTC/CanariCam $8.7\,\mu$m images after applying a correction for point
source emission. We found a good agreement between them to within
$15-20\%$.  We  
also extracted spectra for all the galaxies as extended sources using
an aperture of $0.52\,{\rm arcsec} \times 2\,{\rm arcsec}$. For NGC~6240N and
IRAS~17208$-$0014  we 
extracted the nuclear spectra as extended sources 
with a $0.52\,{\rm arcsec} \times 1\,{\rm arcsec}$ extraction
aperture. 

In Figs~\ref{fig:Mrk1066_images} through \ref{fig:IRAS17208_images}
(bottom panels)
we compare the GTC/CanariCam flux-calibrated nuclear  and 
$0.52\,{\rm arcsec} \times 2\,{\rm arcsec}$ spectra with the 
  CASSIS
{\it  Spitzer}/IRS SL 
spectra. The SL IRS spectra have an approximate angular
  resolution of  $\sim 4\,$arcsec. 

\begin{table*}
 \centering
 \begin{minipage}{155mm}
  \caption{Measurements of the $11.3\,\mu$m
    PAH feature.}\label{table:PAHfeature}  

  \begin{tabular}{lcccccccc}
 \hline
 Galaxy     & \multicolumn{3}{c}{GTC/CanariCam Nuclear } &
 \multicolumn{3}{c}{{\it Spitzer}/IRS Circumnuclear } & $L_{11.3\mu{\rm m \, PAH}}$\\ 
            & Region & EW & Lum& Region & EW & Lum & Nuc/Circum\\
            &(pc)  & ($\mu$m)     & (erg s$^{-1}$) &   (kpc) &  ($\mu$m) & (erg s$^{-1}$)\\ 
\hline
Mrk~1066    & $\le 116$ & $0.35\pm0.02$ &  $5.6\times 10^{41}$ 
& 0.8 & $0.53\pm 0.01$ &   $1.6\times 10^{42}$ & 0.35\\
Mrk~1073    & $\le 230$ & $0.08\pm0.03$ &  $3.2\times 10^{41}$ 
& 1.6 & $0.43\pm 0.01$ &   $2.8\times 10^{42}$ & 0.11\\
NGC~2273    & $\le 64$ & $0.03 \pm 0.01$ &  $2.8\times 10^{40}$
&0.5 & $0.33\pm 0.01$ &   $3.1\times 10^{41}$ & 0.09\\
NGC~3690 & $\le 111$ & $\le 0.01$ & $\le 9.3\times 10^{40}$ 
& 0.8 & $0.13\pm0.01$ & $1.6\times10^{42}$ & $\le 0.06$\\
IC~694 & $\le 111$ & $0.14\pm 0.01$ & $2.9\times 10^{41}$ 
& 0.8 & $0.51\pm0.01$ & $2.5\times 10^{42}$ & 0.12\\
NGC~6240N   & $247 \times 475$& $0.86\pm0.11$ &  $1.4\times 10^{42}$
&1.8 & $0.62\pm 0.01$$^{*}$ &   $7.1\times 10^{42}$$^{*}$ & 0.20\\
NGC~6240S   & $\le 247$ & $0.27\pm0.01$ &  $2.0\times 10^{42}$ & 
 &  & & 0.28 \\
IRAS~17208$-$0014 & $421\times 809$ & $0.56 \pm 0.08$ &  $2.4\times
10^{42}$ &3.0  & $0.90\pm 0.01$ &   $7.8\times 10^{42}$ & 0.31\\
\hline
\end{tabular}

Note.--- For the nuclei not resolved by CanariCam the quoted size (diameter)
of the region is less or equal 
to the slit width. For the {\it Spitzer}/IRS SL data the size of the
region is the slit 
width. We applied a multiplicative factor of 2 to the measured  $11.3\,\mu$m PAH luminosities to be able to compare with results obtained with
{\sc pahfit}. See Section~4.3 for details. $^{*}$The {\it Spitzer}/IRS SL
spectroscopy of NGC~6240 includes the two nuclei of the system (Armus
et al. 2006).
\end{minipage}
\end{table*}

The 2D GTC/CanariCam spectra of Mrk~1066, Mrk~1073, and NGC~6240 show
high S/N
extended $11.3\,\mu$m PAH emission and continuum emission
(see Figs~1, 2, and 4,
respectively). Therefore, for these galaxies we investigate the
spatial variations of the EW and 
flux of the  $11.3\,\mu$m PAH feature and local continuum in
Section~\ref{sec:spatial_variations}.  
To do so, we  used {\sc redcan} to extract from the wavelength
calibrated 2D images  the 
nuclear spectra as well as spectra at  
different radial distances along the slit out to $1.5-2$\,arcsec  on
both sides of  
the nuclei by binning together four pixels (0.32\,arcsec) in the
spatial direction and assuming an extended source extraction.

\subsection{Measuring the $11.3\,\mu$m PAH feature}\label{sec:pahmeasure}
For each of the extracted GTC/CanariCam and {\it Spitzer}/IRS 
spectra we measured the flux and EW  of the $11.3\,\mu$m PAH feature following
the method described in \cite{HernanCaballero2011} and implemented for 
ground-based spectroscopy by \cite{Esquej2014}. We fitted a local 
continuum at $11.25\,\mu$m by interpolating between two narrow bands
($10.75-11.0\,\mu$m and 
$11.65-11.9\,\mu$m) on both sides of the feature. To measure the flux
of the PAH feature we integrated the 
spectrum in the range $\lambda_{\rm rest} =  
11.05-11.55\,\mu$m \citep[see][for full details]{HernanCaballero2011}.  

We estimated 
the uncertainties in the measurements by performing Monte
Carlo simulations using the calculated dispersion
around the measured fluxes and EWs in a hundred simulations
of the original spectrum with random noise distributed as
the rms for the GTC/CanariCam data and error of the spectra for the
Spitzer/IRS data. PAH fluxes obtained with a local continuum are
lower than those using a continuum fitted over a
large spectral range \citep[e.g. measurements done with {\sc pahfit},
see ][]{Smith2007}. We therefore corrected the measured fluxes of the
$11.3\,\mu$m PAH feature  by
applying  a multiplicative factor 
of two as derived by \cite{Smith2007}.  For more details we refer the
reader to \cite{Esquej2014}. We list the $11.3\,\mu$m PAH measurements
in Table~\ref{table:PAHfeature}, including those of the two nuclei of
Arp~299  (NGC~3690 and IC~694).

For reference, galaxies with 
${\rm EW}(11.3\,\mu{\rm m\, PAH})\le 0.1\,\mu$m are AGN
dominated in the mid-IR, whereas those with ${\rm EW}(11.3\,\mu{\rm m\, PAH})\ge
0.2\,\mu$m would have a starburst contribution to the mid-IR emission
greater than 50\% \citep[see][]{HernanCaballero2011}.

\section{Extended nuclear mid-IR emission}\label{sec:spatial_variations}
In this section we present the results on the extended
nuclear mid-IR 
emission as derived from the GTC/CanariCam imaging and spectroscopy
 and compare them with {\it Spitzer}/IRS SL
spectroscopy, which probes larger physical scales. 

\subsection{Mrk~1066}
The GTC/CanariCam $8.7\,\mu$m image of Mrk~1066
(Fig.~\ref{fig:Mrk1066_images}) shows a bright unresolved nuclear 
source (FWHM$\le 67\,$pc, see Table~\ref{table:aperture_photometry})
as well as disk-like extended emission detected out to radial distances of
approximately 1.5\,arcsec. The extended component has  a morphology
similar to that  
observed in the near-IR continuum \citep{Quillen1999} and  Pa$\beta$
and Br$\gamma$ hydrogen recombination lines 
\citep[see Figure~4 of][]{Riffel2010}. The region with extended
emission contains
the narrow line region (NLR) detected in the optical [O\,{\sc
  iii}]$\lambda 5007$ emission line and also regions with active star
formation \citep[see][and references therein]{Riffel2010}.
The similarity between  
the extended $8.7\,\mu$m
emission\footnote{The GTC/CanariCam Si-2 filter includes contributions
from both the $8.6\,\mu$m PAH feature and the underlying continuum, as can be
seen from Fig.~4 in \cite{DiazSantos2008}.} and hydrogen
recombination lines is common in other IR bright galaxies and star
forming galaxies \citep{Helou2004,AlonsoHerrero2006,DiazSantos2008}. This in
principle suggests that the mid-IR  extended emission at $8\,\mu$m 
traces sites of on-going or recent star formation activity but there
can be some contribution from dust in the NLR in AGN \citep[see
e.g.][]{Radomski2003,Packham2005}.

For the GTC/CanariCam spectroscopy, we
oriented the slit along the major axis of the 
extended mid-IR emission  at a PA$=315^{\circ}$. In Fig.~\ref{fig:Mrk1066_images} we
compare the nuclear  and $0.52\,{\rm arcsec} \times
2\,{\rm arcsec}$ GTC/CanariCam spectra with the {\it Spitzer}/IRS
spectrum. The nuclear spectrum probes approximately a region of
116\,pc in size\footnote{Since this galaxy is inclined at $\sim
  50^{\circ}$, the region encompassed by the slit is larger than that
  indicated by the projected slit width.} and  shows 
bright $11.3\,\mu$m PAH emission (also from the $8.6\,\mu$m PAH
feature). This is indicative of the presence of nuclear star
formation activity. There is also evidence of the  
silicate feature in absorption. The nuclear contribution
to the $\sim 4\,$arcsec ($\sim 0.9\,$kpc) $11.3\,\mu$m PAH feature
measured in the {\it Spitzer}/IRS spectrum 
 is 35\% (see Table~\ref{table:PAHfeature}).  Based
on the EW of the
$11.3\,\mu$m PAH feature in the {\it Spitzer}/IRS spectrum
\cite{HernanCaballero2011} classified this galaxy as starburst, while
the nuclear value of the EW is lower, as expected, due to the higher
contribution from the AGN continuum  (see discussion in
Section~\ref{sec:EWPAH}, and also Ramos Almeida et al. 2014).

\begin{figure}

   \resizebox{0.96\hsize}{!}{\rotatebox[]{0}{\includegraphics{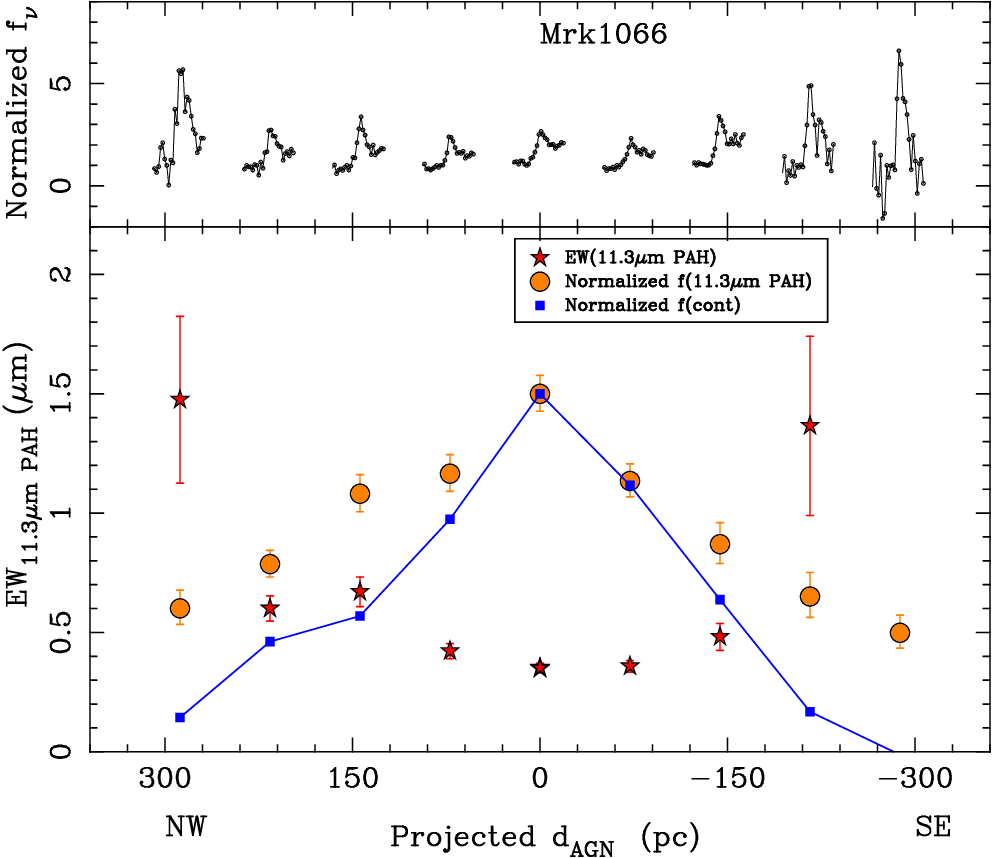}}}
\vspace{0.25cm}

   \caption{Lower panel. GTC/CanariCam spatial profiles of Mrk~1066
     as a function of the projected distance from the AGN
     $d_{\rm AGN}$. We show the profiles of the  EW and flux
      of the $11.3\,\mu$m PAH feature as star-like symbols and filled
      dots, respectively, and the flux of the local continuum at
     $11.25\,\mu$m as filled
     squares. The measured 
     fluxes are normalized to the peak intensity at the nuclear
     position, which is set at the 0 position on the x-axis, and set
     to a 1.5 value on the y-axis. For clarity we plotted only the errors in the
     measured EW and flux of the PAH feature. Top panel.
     Spectra  around the $11.3\,\mu$m PAH feature
     ($\lambda_{\rm rest}=10.5-12\,\mu$m), normalized to the same
     continuum level at $\sim 10.9\,\mu$m for the
     spatial location of each of the extracted 
     spectra.}\label{fig:spatial_variationsMrk1066}  
\end{figure}

We can make use of the
spatial information afforded by the GTC/CanariCam spectroscopy to
produce spatial profiles of the mid-IR emission
(Section~4.2). Fig.~\ref{fig:spatial_variationsMrk1066}   
shows the spatial profiles of the EW
of the $11.3\,\mu$m PAH feature (star symbols) and its normalized flux
(circles) together with the normalized 
local continuum (squares) as a function of the projected distance 
from the AGN, $d_{\rm AGN}$. The $11.3\,\mu$m PAH
emission is detected out to projected radial distances 
of $\sim 300\,$pc on both sides of the nucleus. The
intensity of the PAH feature  peaks
at the nucleus. From the modelling of the optical spectrum
\cite{GonzalezDelgado2001} showed that 
the nuclear emission of Mrk~1066 is dominated by 
the emission of young and intermediate age stellar populations. Since we
detected nuclear $11.3\,\mu$m PAH
emission in the central $\sim 100\,$pc and  PAH
molecules are excited by the UV emission from O stars and B stars
\citep{Peeters2004}, we conclude that the $11.3\,\mu$m PAH emission 
can be used to reveal the presence of on-going/recent nuclear star 
formation in the nuclear regions of AGN.

The spatial profile of the local interpolated continuum at
$11.25\,\mu$m measured from the GTC/CanariCam spectrum has 
its maximum also at the nucleus, but it is narrower 
than that of the PAH feature. This indicates that in the nuclear
region the continuum at this wavelength has a strong component
presumably arising from
the unresolved AGN emission. Additionally, the  PAH
spatial profile
shows a secondary peak at a radial distance of $\sim 150\,$pc
to the northwest that is approximately
coincident with one of the starburst  sites detected in the near-IR
\citep[see][]{Riffel2010}.  Also, on both sides of the nucleus the PAH
flux profile extends out to the radial distances of $\sim 300\,$pc.

The spatial
profile of the EW of the $11.3\,\mu$m PAH feature shows a minimum
(EW$=0.35\pm0.02\,\mu$m) at the nuclear position, which is typical of 
 Seyfert galaxies with nuclear/circumnuclear star formation
\citep{Sales2010, Tommasin2010,
    HernanCaballero2011,Esquej2014}. At larger   
radial distances from the nucleus, the EW of the feature increases
reaching values typical of star forming galaxies (EW$\sim 0.5-1\,\mu$m).  This
behaviour is well understood in terms of the increased   continuum from the AGN 
towards the nuclear regions rather than a fall-off of the PAH
emission because the $11.3\,\mu$m PAH feature peaks at the nucleus (see
a detailed discussion in Section~\ref{sec:EWPAH}). Even
at the location of the two starburst sites, the observed EW of the feature 
still has a relatively low value due to contamination from continuum
produced by AGN heated
dust. Indeed, when the extracted spectra are corrected
for this AGN contamination, the measured EW are consistent with pure
star formation. We refer the reader to \cite{RamosAlmeida2014} for a detailed study of
the star formation and torus properties of this galaxy.

\subsection{Mrk~1073}
The GTC/CanariCam $8.7\,\mu$m image (Fig.~\ref{fig:Mrk1073_images})
shows a bright unresolved nuclear source with a FWHM size of $\le 146\,$pc
(see Table~\ref{table:aperture_photometry}). There is also extended
emission in the disk of the galaxy with an approximate diameter of
$2.6\,{\rm arcsec} \simeq 1.1$\,kpc.  For the spectroscopy we placed the
GTC/CanariCam slit along the major axis of this emission (PA$=75^{\circ}$).  

The GTC/CanariCam nuclear 
spectrum  of Mrk~1073 shows the silicate feature in moderate absorption with
weak $8.6\,\mu$m  and $11.3\,\mu$m PAH emission
(Fig.~\ref{fig:Mrk1073_images}, bottom panel). The PAH features
become more apparent on the $\sim 1.6\,$kpc scale probed by the {\it
  Spitzer}/IRS spectrum with an EW of the $11.3\,\mu$m PAH feature
being consistent with a composite AGN/starburst nature 
  \citep{Sales2010, Tommasin2010, 
    HernanCaballero2011, Esquej2014}.  
The nuclear contribution to the 
observed $11.3\,\mu$m PAH feature luminosity in the central 1.6\,kpc
of Mrk~1073 is only $11\%$, but consistent with findings for other 
Seyfert galaxies \citep[see][]{Esquej2014}. The low value of the
nuclear EW of the $11.3\,\mu$m PAH feature (EW=$0.08\pm0.03\,\mu$m, see 
Table~\ref{table:PAHfeature}) clearly
indicates that on nuclear scales the mid-IR emission of Mrk~1073 is
dominated by emission from the AGN, unless the PAHs get destroyed by
the AGN (but see Section~6.3).

 \begin{figure}

   \resizebox{0.96\hsize}{!}{\rotatebox[]{0}{\includegraphics{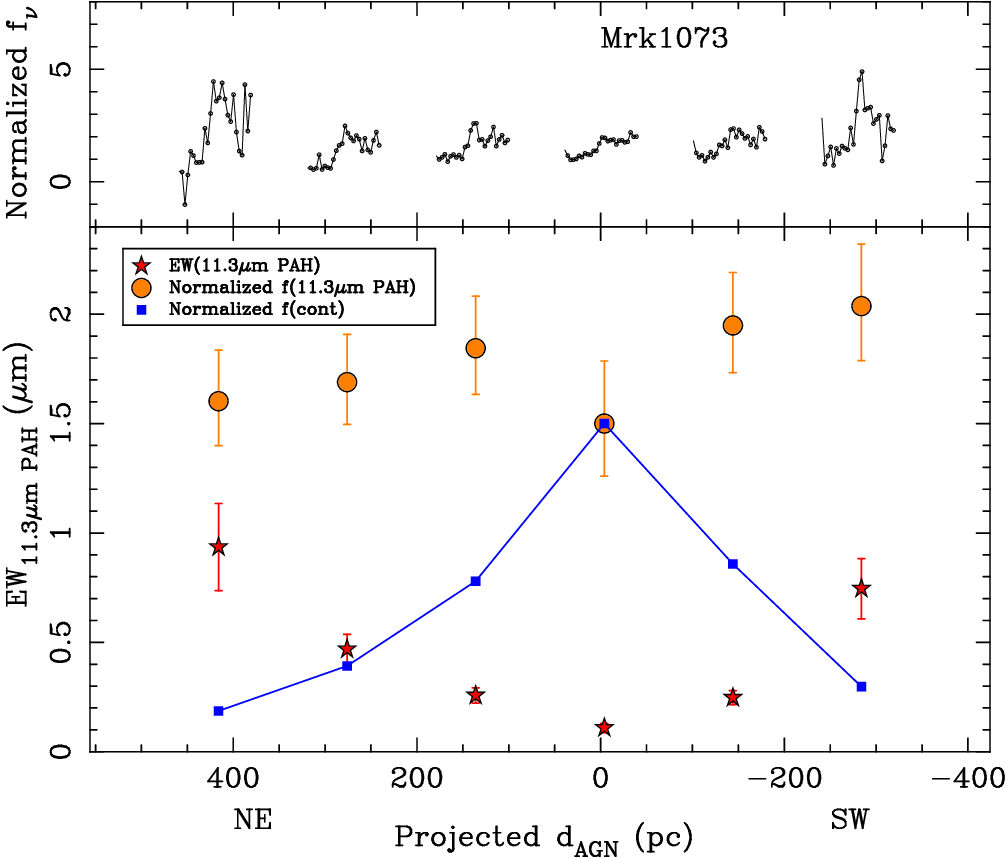}}}
\vspace{0.25cm}

   \caption{As Fig.~\ref{fig:spatial_variationsMrk1066}, but for
     Mrk~1073. }\label{fig:spatial_variationsMrk1073}
\end{figure}

 \begin{figure*}

   \resizebox{0.7\hsize}{!}{\rotatebox[]{0}{\includegraphics{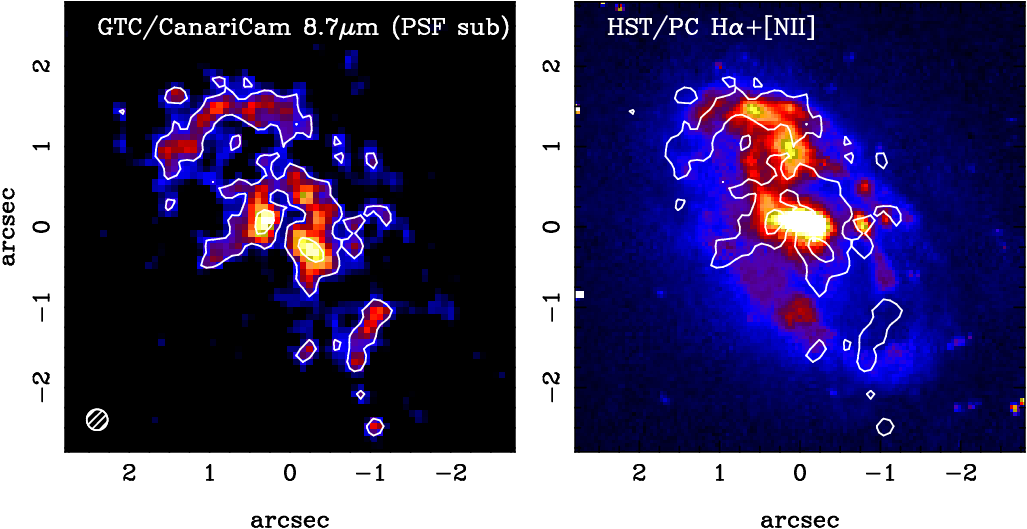}}}
   \caption{NGC~2273. Left panel. GTC/CanariCam PSF-subtracted
     $8.7\,\mu$m in a square-root colour scale. The hatched circle 
is the FWHM of the PSF. Right panel. 
{\it HST}/PC H$\alpha$+[N\,{\sc
       ii}]+continuum image with the $8.7\,\mu$m contours
     superimposed. The lowest $8.7\,\mu$m contour is as in
     Fig.~\ref{fig:NGC2273_images} (top panel).
 }\label{fig:midIRopticalNGC2273}    

\end{figure*}

Fig.~\ref{fig:spatial_variationsMrk1073} shows the spatial profiles of
the flux and EW of the $11.3\,\mu$m PAH
feature emission and the flux of the local continuum at
$11.25\,\mu$m. The PAH feature  
is detected out to projected distances from the AGN of $\sim 400\,$pc
and $\sim 300\,$pc to the northeast and southwest of the nucleus,
respectively. This 
confirms that the extended $8.7\,\mu$m emission detected in the GTC/CanariCam
image is due to star formation activity. Unlike
Mrk~1066, the PAH feature emission does not peak strongly at the nucleus, but
rather it is approximately uniform along the
spatial direction at PA$=75^{\circ}$. This is in good agreement with the
relatively 
constant contribution of the
young stellar populations ($\le 100\,$Myr) in the central few kpc of
this galaxy \citep{Raimann2003}. We note that the spatial profiles
plot the observed (not corrected for extinction) flux of the
$11.3\,\mu$m PAH feature. The slight decrease of the flux of
the PAH feature at the location of the nucleus might be due a higher
extinction there as indicated by the moderate depth of the silicate feature. 
The EW of the $11.3\,\mu$m on the
other hand, has its minimum value in the nuclear region, consistent
with the presence of a strong AGN produced continuum there. At
projected distances $\sim 300\,$pc and further away from the nucleus
the measured 
values of the EW are consistent with that observed 
in regions of pure star formation activity.

\subsection{NGC~2273}
The GTC/CanariCam $8.7\,\mu$m image of NGC~2273 (see
Fig.~\ref{fig:NGC2273_images})  
shows a bright nuclear point source with an unresolved size 
of $\le 37\,$pc (FWHM). There is also diffuse emission out to  a radial
distance of approximately $1.5\,{\rm arcsec}=190\,$pc, which appears to be more
prominent to the north of the nucleus. 

\cite{Ferruit2000} detected  a
ring of star formation using {\it HST} narrow-band H$\alpha$+[N\,{\sc
  ii}] imaging. To compare the extended mid-IR and optical 
emission, we first 
subtracted the bright nuclear unresolved component in the
GTC/CanariCam images. We used the
standard star observation taken for the photometric calibration as a
representation of the PSF and followed the method for PSF subtraction
explained by 
\cite{RamosAlmeida2009}. We note that we did the PSF subtraction 
before rotating and smoothing the NGC~2273 image. We used the bright
nuclear emission to align the optical and mid-IR images.

 \begin{figure*}

   \resizebox{0.7\hsize}{!}{\rotatebox[]{0}{\includegraphics{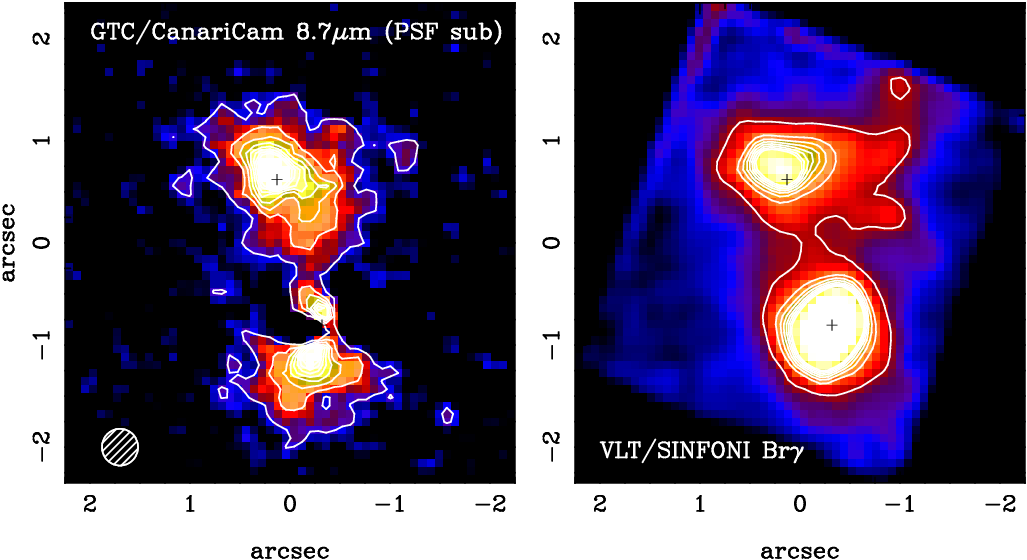}}}
   \caption{NGC~6240. Left panel. GTC/CanariCam PSF-subtracted
     $8.7\,\mu$m (only unresolved emission from the southern nucleus
     was subtracted)
     in a square-root colour scale. The hatched circle 
is the FWHM of the PSF.  The contours are the $8.7\,\mu$m emission
  with the lowest value as in Fig.~\ref{fig:NGC6240_images} (top panel). Right
panel. VLT/SINFONI 
     continuum-subtracted 
     Br$\gamma$ image in a square-root colour scale from Engel et
     al. (2010). We  
     repixelated and smoothed the SINFONI image to match the pixel size and
     angular resolution of the  
     CanariCam image. The contours are the Br$\gamma$ emission
  with the lowest value set at the background level plus 8 standard
  deviations (before smoothing the image). }\label{fig:midIRSINFONINGC6240}   

\end{figure*}

Fig.~\ref{fig:midIRopticalNGC2273} shows the PSF-subtracted
$8.7\,\mu$m image compared with the H$\alpha$+[N\,{\sc ii}] image. The
diffuse emission  detected at 
$8.7\,\mu$m to the north of the nucleus at $\sim 1.5\,$arcsec has a 
reasonable resemblance with the optical line emission. Some of the
differences might be due to the fact that we did not subtract the
continuum from the optical line image. Therefore, there is  an additional
component due to stellar emission. To the south  
of the nucleus the emission from the ring appears less prominent. In
the central $2\,{\rm arcsec} = 250\,$pc, the PSF-subtracted $8.7\,\mu$m image 
reveals the presence of extended nuclear emission\footnote{We note
  that some of the structure seen there,
  especially the two knots at PA of about 50$^{\circ}$, might be due to
imperfect PSF subtraction. } with a similar
orientation as the H$\alpha$+[N\,{\sc ii}] nuclear emission. Since
there is also diffuse extended $11.3\,\mu$m PAH 
emission in the nuclear region, as can be seen from the 2D GTC/CanariCam
spectrum (Fig.~\ref{fig:NGC2273_images}), 
the extended emission in the Si-2 filter has likely a 
contribution from the $8.6\,\mu$m PAH feature. This extended nuclear
emission might be related to dense molecular gas emission
detected in the nuclear region (central $\sim 2\,$arcsec) 
of this galaxy \citep[see ][]{Sani2012} and the nuclear ring 
identified by \cite{Barbosa2006} by the presence of a velocity
dispersion ($\sigma$) drop. The nuclear {\it HST} optical line ratios
appear to have some contribution  from the NLR
emission \citep{Ferruit2000}.

The GTC/CanariCam nuclear spectrum of NGC~2273 (see
Fig.~\ref{fig:NGC2273_images}) is dominated by the
bright AGN continuum with weak, but detectable, $11.3\,\mu$m PAH  
emission with EW=$0.03\pm 0.01\,\mu$m.  
The modest angular resolution and high sensitivity of the {\it
  Spitzer}/IRS SL spectrum, on the other hand capture the diffuse
emission within the central $4\,{\rm arcsec} \sim 500\,$pc as
bright $8.6\,\mu$m and $11.3\,\mu$m PAH emission. The measured EW of
the $11.3\,\mu$m PAH feature within the IRS aperture reflects the starburst/AGN composite nature of
the emission 
  \citep[see][]{Sales2010, Tommasin2010, 
    HernanCaballero2011, Esquej2014}.   

The nuclear
contribution to the observed $11.3\,\mu$m PAH emission within this
region is 9\%, thus indicating that most of the star formation in the
central 500\,pc of this galaxy arises from the circumnuclear
ring (that is, outside the inner $\sim 60\,$pc). Still, the detection
of the $11.3\,\mu$m PAH 
feature in the nuclear region (inner
$\le 64$\,pc) implies a
star formation rate of $0.08\,M_\odot \, {\rm yr}^{-1}$
($0.21\,M_\odot \, {\rm yr}^{-1}$ for the $\sim 60\,{\rm pc} \times
250\,{\rm pc}$
aperture),  
using the calibration of \cite{DiamondStanic2012}. This appears  in
contradiction with the Toomre parameter calculated by \cite{Sani2012}
that seems to indicate that
the central $\sim 100\,$pc of this galaxy should be
stable against star formation.

\subsection{Arp~299}\label{sec:Arp299}
The GTC/CanariCam observations and modeling of the AGN properties
of this interacting system were 
presented in \cite{AlonsoHerrero2013}. Here we only discuss briefly 
the $11.3\,\mu$m PAH feature. In the nuclear region of the eastern 
component of  Arp~299, IC~694,
the nuclear $11.3\,\mu$m PAH feature is detected with EW$=0.14 \pm
0.01\,\mu$m, which is indicative of a composite starburst/AGN nature. 
\cite{AlonsoHerrero2000} concluded that the IR emission of
this component is dominated by star formation activity. However, there
is also evidence of the presence of AGN activity, based on the
reduced EW when compared to that measured on kpc scales with
{\it Spitzer}/IRS and the modelling of the nuclear spectrum
\citep{AlonsoHerrero2013}.

 \begin{figure}

   \resizebox{0.96\hsize}{!}{\rotatebox[]{0}{\includegraphics{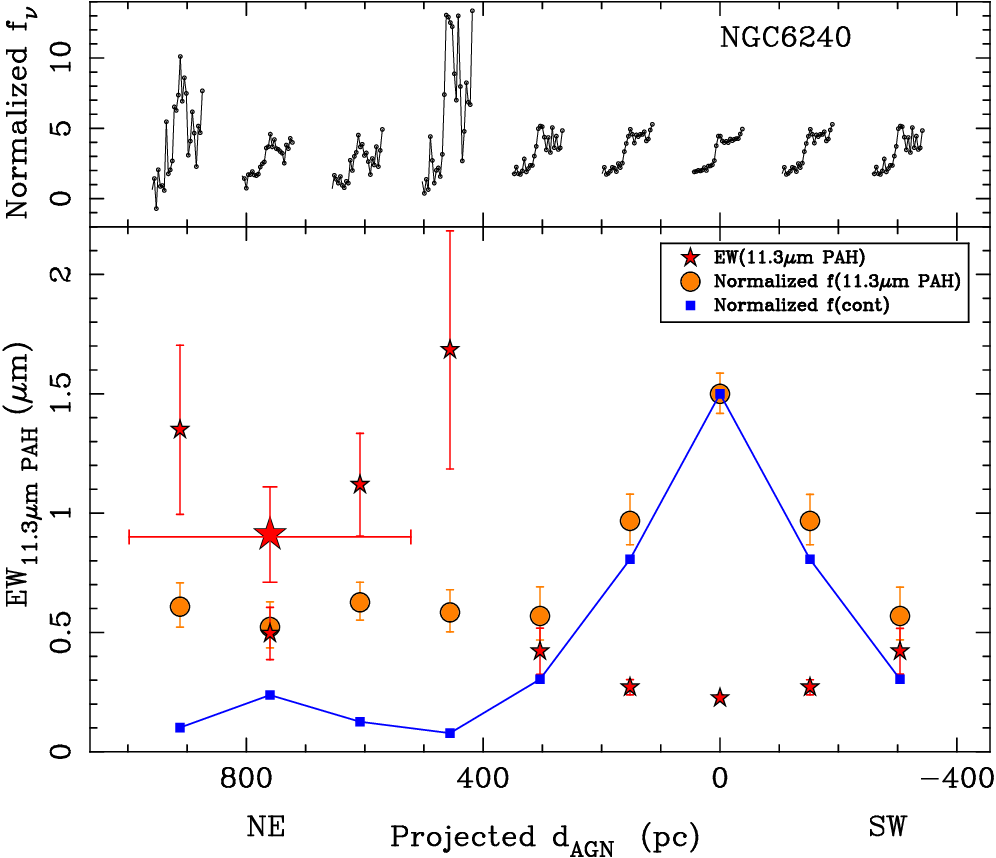}}}
\vspace{0.25cm}

   \caption{As Fig.~\ref{fig:spatial_variationsMrk1066}, but for
     NGC~6240. The 0 position on the x-axis marks the location of the
     southern nucleus. The secondary peak in the local continuum at
     $11.25\,\mu$m at a projected distance of 760\,pc to
     the northeast marks the approximate location 
     of the northern nucleus of
     NGC~6240. The large star symbol is the measurement of the EW in
     the northern nucleus from the $0.52\,{\rm arcsec} \times 1\,{\rm
       arcsec}$ 
     spectrum (see also Table~\ref{table:PAHfeature}), shown as the
     horizontal error 
     bar. }\label{fig:spatial_variationsNGC6240}    
\end{figure}

In the case of NGC~3690, the nucleus of the western component of 
 Arp~299, the nuclear (distance from the AGN $d_{\rm AGN}\le 50\,$pc)
$11.3\,\mu$m PAH feature is 
not detected. The $2\sigma$ upper limit is ${\rm EW}\le
0.01\,\mu$m, whereas in the IRS spectrum we measured
${\rm EW}=0.13\pm0.01\,\mu$m. \cite{AlonsoHerrero2000} showed that the 
star formation rate in the nucleus of NGC~3690 is about five time less
than in IC~694.  The upper
limit (at the $2\sigma$ level) to the $11.3\,\mu$m PAH luminosity in
the nucleus of NGC~3690 compared with that of IC~694 (see
Table~\ref{table:PAHfeature}) is compatible with this. We cannot 
therefore reach a firm conclusion as to whether PAHs are destroyed
near the nucleus of NGC~3690 or the low EW is due to dilution by
the luminous AGN continuum (Section~6.1). We also note that 
the {\it HST}/NICMOS Pa$\alpha$ image shows that there is extended
emission approximately 1.1\,arcsec southwest from the nucleus of
NGC~3690 \citep[i.e. sources B16 and B1, respectively, 
see figures~2b and 7b in][]{AlonsoHerrero2000}, which
is likely to be associated with the star formation detected in the
{\it Spitzer}/IRS spectrum. These regions of star formation, however,
did not fall in the GTC/CanariCam slit. 

\subsection{NGC~6240}
NGC~6240 was previously imaged in the mid-IR using MIRLIN
on the Keck II telescope \citep{Egami2006} and VISIR on the VLT
telescope \citep{Asmus2014}. The GTC/CanariCam $8.7\,\mu$m image of
NGC~6240 (Fig.~\ref{fig:NGC6240_images}, top panel) shows 
mid-IR emission arising from both nuclei, 
with the southern nucleus
being approximately five times brighter than 
the northern one (Table~\ref{table:aperture_photometry}). The
CanariCam image shows also
diffuse emission around and in between the nuclei. We detect extended
emission in the central $4\,{\rm arcsec} \sim 1.9\,$kpc of the system. The
core of the southern
nucleus appears unresolved at the resolution of the CanariCam images
($\le 181\,$pc, FWHM), while the northern nucleus is resolved
\citep[see also][]{Asmus2014} with a ${\rm FWHM}=214$\,pc. 

In order to study the extended mid-IR $8.7\,\mu$m  emission in more detail we
subtracted the unresolved emission (PSF subtraction) from the southern
nucleus of NGC~6240. We show the result in Fig.~\ref{fig:midIRSINFONINGC6240}
compared with the VLT/SINFONI continuum-subtracted Br$\gamma$ image from
\cite{Engel2010}. The latter emission in NGC~6240 is mostly believed
to be tracing the starburst activity in this galaxy. We repixelated
and smoothed the SINFONI image to match the pixel size and angular
resolution of the CanariCam image.

As can be seen from this figure,
there is a fairly good correspondence between both emissions,
especially around the northern nucleus where the AGN contribution in
the mid-IR is
low (see below). There is also extended
$8.7\,\mu$m emission to the south of the southern nucleus, which has also been
detected in the near-IR continuum, Br$\gamma$, and other mid-IR
wavelengths \citep{Asmus2014}. This again indicates that the
extended $8.7\,\mu$m emission probes recent star formation activity. 
Some of the differences in morphologies might be attributed to the
high extinction affecting the $K$-band emission in and around 
the nuclei (see Figure~9 in Engel et al. 2009). It is also interesting
to note that the peaks of the Br$\gamma$ line emission and the
$8.7\,\mu$m emission around the northern nuclei are nearly coincident
with the northern radio source \citep{Beswick2001} to within one 
CanariCam pixel
($\sim 0.08\,$arcsec), while that is not the case for the peak of the 
$2\,\mu$m continuum \citep[see][]{Max2007,Engel2010}. The radio
source is believed to mark the position of the accreting black hole in
the northern nucleus. At $\lambda > 2\,\mu$m
the peaks coincide with the radio ones \citep{Mori2014}.

Fig.~\ref{fig:NGC6240_images} (bottom panel) compares the CanariCam
nuclear spectra with that  
observed through the {\it Spitzer}/IRS SL $\sim 4\,$arcsec aperture, which
contains the two nuclei \citep{Armus2006}. Both nuclear spectra  show
relatively bright 
emission from the 8.6 and $11.3\,\mu$m PAH features. The IRS spectrum
and that of the southern nucleus also show bright molecular hydrogen
lines H$_2$ at $8.025$, $9.665$, and $12.279\,\mu$m.
The spatial profiles of Fig.~\ref{fig:spatial_variationsNGC6240} show
that there is $11.3\,\mu$m PAH emission extending $3\,{\rm arcsec} \sim
1.4\,$kpc along
the slit. Both the local continuum at $11.25\,\mu$m and the $11.3\,\mu$m PAH feature 
peak at the southern nucleus, although the PAH emission appears to be
slightly more extended. There is also extended PAH
emission to the southwest of this 
nucleus where there is also detected extended emission in the
$8.7\,\mu$m and Br$\gamma$ images. 
Around the northern nucleus the PAH emission is fainter than in the
southern nucleus but it 
appears rather constant extending approximately 1.5\,arcsec ($\sim
700\,$pc) including the region between the two nuclei. 

The spatial 
profile of the EW of the PAH feature shows a minimum
(EW=$0.28\pm0.02\,\mu$m) at the southern nucleus and a secondary
minimum at the location of the northern nucleus (EW=$0.50\pm
0.11\,\mu$m). This clearly  
indicates the presence of a continuum, most likely associated
with dust heated by the AGN in both nuclei, which is more dominant in
the southern nucleus \citep[see][for a detailed
analysis of the AGN IR emission of the southern nucleus]{Mori2014}. To the
southwest of the southern nucleus and in the regions around the northern nucleus
the
EW of the feature is similar to that measured in starburst galaxies,
as is also the case for the {\it Spitzer}/IRS spectrum.

\subsection{IRAS~17208$-$0014}
The GTC/CanariCam $8.7\,\mu$m
image of IRAS~17208$-$0014 (Fig.~\ref{fig:IRAS17208_images})  shows
emission extending over 
approximately 2\,arcsec $= 1.6\,$kpc. The morphology is similar to the Keck
mid-IR 7 and $12\,\mu$m images
\citep[see][]{Soifer2000}  and the 
VLT/SINFONI Pa$\alpha$ emission
\citep{PiquerasLopez2012}.  The measured size of the 
nuclear region at $8.7\,\mu$m, which appears resolved in the CanariCam
image, is $0.52\,{\rm arcsec}=421$\,pc (FWHM), and therefore there is no
clear evidence of the putative nuclear  
unresolved emission expected to be associated with the AGN. 

The 2D GTC/CanariCam spectrum (see Fig.~\ref{fig:IRAS17208_images})
shows extended PAH emission.  
The measured nuclear EW of the $11.3\,\mu$m feature 
(EW=$0.56\pm 0.08\,\mu$m) shows that on scales of a few hundred
pc the mid-IR emission of this galaxy is dominated by
star formation.  However, the lower
nuclear value of the EW of the $11.3\,\mu$m PAH feature  compared to
the value measured in the IRS spectrum  (EW=$0.90\pm
0.01\,\mu$m) indicates a higher continuum contribution there, which
might be due to the AGN-produced mid-IR continuum. This is
similar to findings for the nuclei of Arp~299 (see
Table~\ref{table:PAHfeature}). Alternatively, the
decreased nuclear 
EW in this galaxy could be due to an increased starburst continuum, as
found in some purely star forming nuclei
\citep{TacconiGarman2005,DiazSantos2010}. 
Finally, the nuclear
contribution to the {\it Spitzer}/IRS $11.3\,\mu$m PAH feature
emission in IRAS~17208$-$0014 is 31\%, indicating that the emission of
this feature is 
extended over at least a few kpc.

\section{Discussion}\label{sec:discussion} 
For this work we selected six systems (8 galaxies) known to harbour
an AGN and have intense  star formation
activity on nuclear and circumnuclear regions. We
detected $11.3\,\mu$m PAH feature emission in all the galaxy nuclei 
(inner regions between 64 and 420\,pc), except
NGC~3690. Moreover,  because some of them are 
LIRGs or ULIRGs they tend to show more highly concentrated $11.3\,\mu$m PAH
emission (i.e. star formation rate)  than typical Seyferts
\citep{Esquej2014}. This would be in 
line with the interpretation that local LIRGs may represent an 
IR-bright star-forming phase taking place prior to most of the AGN phase
\citep{AlonsoHerrero2013LIRGs}. The galaxies in this work also have
relatively high X-ray column densities, which may also be related to
the material protecting the PAH molecules in the nuclear regions of
AGN (Section~6.3). In what follows, we discuss in more detail the
nuclear $11.3\,\mu$m 
PAH emission in local AGN.

\subsection{Understanding the EW of the nuclear $11.3\,\mu$m PAH
  feature of Seyfert galaxies}\label{sec:EWPAH}  
The EW of the PAH features is routinely 
used to obtain a general classification of the dominant source of
activity in galaxies (AGN vs. star formation) in the near and mid-IR
\citep[see examples using the 3.3, 6.2, 7.7, and $11.3\,\mu$m PAH features
in][respectively]{Imanishi2000,Spoon2007,Genzel1998,HernanCaballero2011}. It is
also used to detect AGN not previously 
identified in the optical in LIRGs and ULIRGs \citep[see
e.g.][]{Nardini2008,AlonsoHerrero2012}. This is because generally high
metallicity star-forming galaxies show remarkably similar mid-IR
spectra \citep{Brandl2006}, while the nuclear spectra of AGN are
dominated by a strong  continuum due to dust heated by the AGN. This
results in AGN-dominated sources having much 
lower EWs of the PAHs than starburst galaxies. There is also both
observational and theoretical evidence that PAH 
emission can be diminished/suppressed near (scales of a few pc) the
ionizing sources in 
sites of  intense star formation  \cite[e.g.][]{Siebenmorgen2004,
TacconiGarman2005, Tielens2013}.

 \begin{figure}

   \resizebox{0.98\hsize}{!}{\rotatebox[]{-90}{\includegraphics{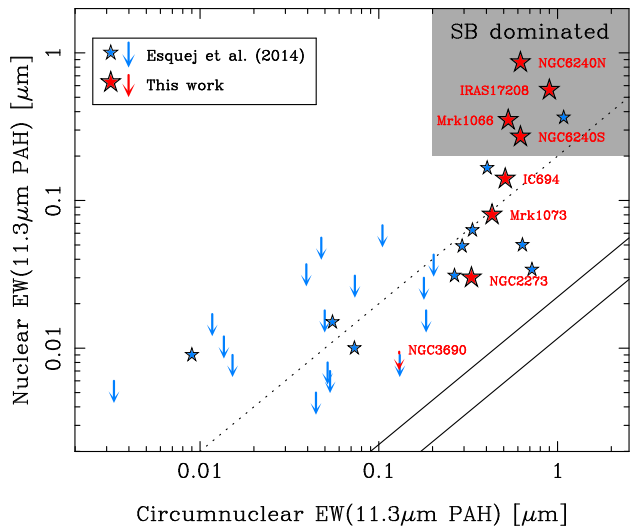}}}
\vspace{-0.5cm}

   \caption{EW of the $11.3\,\mu$m PAH feature
     measured on nuclear scales (this work  red symbols and
     Esquej et al. 2014 blue symbols) from 8-10\,m-class telescopes
     vs. that on circumnuclear kpc scales with {\it
  Spitzer}/IRS SL spectra. The two solid lines
represent the range of expected ratios for the case of uniform PAH emission
taking into account the difference between
ground-based slits ($0.35-0.75\,$arcsec) and the IRS slit width. The dotted
line represents the 
typical 0.2 ratio 
between the nuclear and circumnuclear PAH emission in RSA
Seyferts \citep[see][]{Esquej2014}. The shaded region shows the predicted EW for a starburst contribution $\ge
50\%$ to the observed mid-IR emission
\citep[see][]{HernanCaballero2011}. }\label{fig:comparisonEW}     
\end{figure}

A number of works using mid-IR spectra of AGN 
noted a tendency of decreasing  EWs of PAHs with increasing hardness
of the AGN radiation field 
and interpreted it as evidence of PAH destruction in the nuclear regions of AGN
\citep[see e.g.][]{Wu2009,Sales2010}. However, other interpretations
for the decreased EW of PAH features in AGN include 
dilution from the intense
mid-IR continuum produced by the AGN \citep[see
e.g.][]{Lutz1998,Clavel2000} and  
reduction of the star formation activity from kpc scales to the nuclear
regions of AGN \citep{Hoenig2010}.

Owing to the high angular resolution of the CanariCam data we  showed 
in Section~\ref{sec:spatial_variations} that the 
nuclear EW of the $11.3\,\mu$m PAH feature varies with the 
distance from the AGN,  always increasing at larger radial
distances from the nucleus. This is 
similar to findings for 
Circinus and NGC~1808 using ground-based spatially-resolved mid-IR spectroscopy
\citep[see][respectively]{Roche2006,Sales2013}.
We also demonstrated that since the $11.3\,\mu$m PAH
flux peaks in the nuclear region, the minimum value of its EW 
in the nuclei of these AGN is due to an increased
contribution from the AGN produced continuum and not to PAH
destruction \citep[see also][for the $3.3\,\mu$m PAH feature emission
in the nuclear 
region of Centaurus~A]{TacconiGarman2013}.
Finally, in galaxies with a higher nuclear
concentration of the star formation activity (Mrk~1066, NGC~6240,
and IRAS~17208$-$0014, see last column in  
Table~5), the observed nuclear EW of the PAH feature is larger than
those with smaller nuclear 
fractions of the PAH emission (NGC~2273, Mrk~1073, and NGC~3690). This simply
reflects a higher contribution of the 
star formation activity with respect to the AGN emission. 

\begin{table*}
 \centering
 \begin{minipage}{110mm}
  \caption{EW of the $11.3\,\mu$m
    PAH feature in the spectra of Seyfert nuclei.}\label{table:PAHfeature_composites}  

  \begin{tabular}{lcccccccc}
 \hline
 Class     & No. & \multicolumn{2}{c}{Nuclear } &
 \multicolumn{2}{c}{{\it Spitzer}/IRS } \\ 
            & & Region & EW & Region & EW \\
            & &(pc)  & ($\mu$m)    &   (pc) &  ($\mu$m) \\ 
\hline
Seyfert 1  & 8 & 104 & $0.015\pm0.007$ & 650 & $0.192\pm 0.056$\\
Seyfert 2  & 9 & 104 & $0.028\pm0.016$ & 720 & $0.228\pm0.075$ \\
Moderate absorption & 9 &67 & $0.047\pm0.038$ & 625 & $0.233\pm0.071$\\
\hline
\end{tabular}
Notes.--- No. indicates the number of spectra combined and region
 the median value of the projected slit widths (in pc)
of the nuclear and {\it
  Spitzer}/IRS spectra.
\end{minipage}
\end{table*}

In Fig.~\ref{fig:comparisonEW} we compare the EWs of the $11.3\,\mu$m
PAH feature measured typically on hundreds of pc scales with {\it
  Spitzer}/IRS and nuclear scales with GTC/CanariCam. We also plotted in
this figure data from \cite{Esquej2014} for Seyferts in the revised
Shapley-Ames (RSA) galaxy catalog 
\citep{MaiolinoRieke1995} where the nuclear 
ground-based spectroscopy comes from 8-10\,m telescopes and probes
typically nuclear regions of 60\,pc in size. The solid lines are the
predictions for the nuclear EW in the presence of an AGN 
assuming that the PAH emission is
distributed uniformly taking into account the range of slit sizes of
the nuclear spectra and that of the {\it Spitzer}/IRS spectra. The
dotted line represents the average value 
between nuclear and circumnuclear $11.3\,\mu$m PAH luminosities of 0.2
found for RSA 
Seyferts. Measurements above this line indicate galaxies with strong
nuclear star formation, as also expected from values of the EW if star
formation contributes more than 50\% of the observed mid-IR
emission. Since we selected the galaxies observed with 
CanariCam as having intense nuclear star formation, it is not
surprising to see that the nuclear mid-IR emission in many of them
is dominated by star formation when using
the EW of the feature. It is also
clear from this figure, that 
unless there is a strong nuclear starburst, we should expect that
the nuclear values of the EW are smaller than those measured on
circumnuclear scales from {\it Spitzer}/IRS \citep{Siebenmorgen2007}.

\subsection{Mid-IR spectra of Seyfert nuclei}\label{sec:averagespectra}

In this section we investigate further the detection 
of the $11.3\,\mu$m PAH feature in the close vicinity of AGN 
by constructing mean mid-IR spectra of Seyfert nuclei.
We used the  nuclear (typically the central 65\,pc)
mid-IR ground-based spectra of Seyfert galaxies 
in the RSA sample   compiled by
Esquej et al. (2014, their Table~2), except for the two Seyferts with
  the $10\,\mu$m silicate feature in emission (NGC~2110 and NGC~7213) and
  NGC~1808 whose spectrum is dominated by star formation. 
We  included the three
galaxies (Mrk~1066, Mrk~1073, and NGC~2273) in our sample that are
also in the RSA sample.

We divided the nuclear spectra in three categories, namely, Seyfert 1,
Seyfert 2, and Seyfert nuclei with silicate features in moderate absorption. 
 We chose this last category because in these Seyfert 2 nuclei  
the silicate feature is likely to have a contribution from  extended dust
absorbing components in the host galaxy in addition to that due to 
the dusty torus \citep{Deo2009,AlonsoHerrero2011,Goulding2012,
  GonzalezMartin2013}. These 
nuclei are generally in merger galaxies, galaxies with nuclear dust
lanes, and/or highly inclined systems \citep[see also][]{Hoenig2014}.
We combined the nuclear mid-IR spectra after normalising them at
$12\,\mu$m. 

\begin{figure}

   \resizebox{0.99\hsize}{!}{\rotatebox[]{-90}{\includegraphics{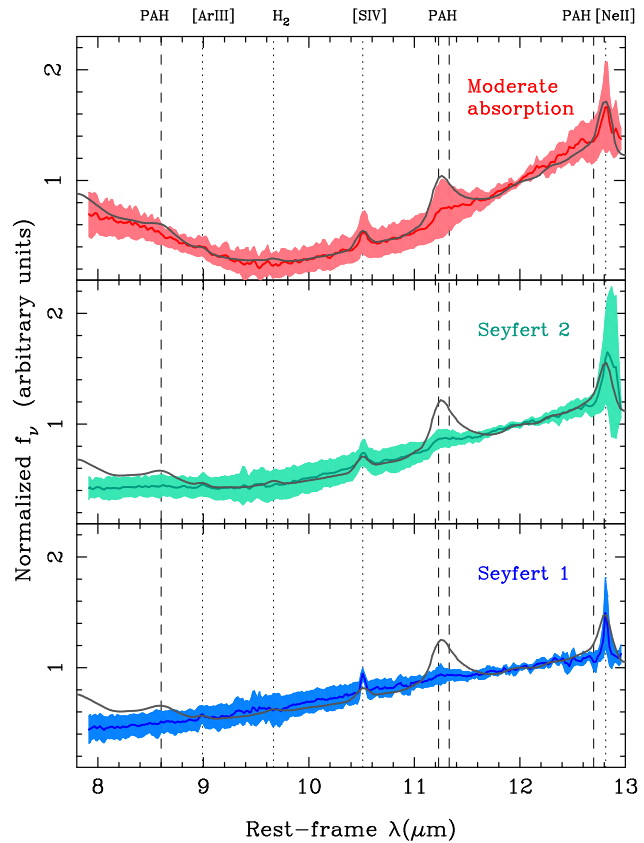}}}

   \caption{Mid-IR spectra
     (normalized at $12\,\mu$m) 
     of Seyfert 1 nuclei (bottom panel, blue line), Seyfert 2 nuclei
      (middle panel, green line), 
     and Seyfert nuclei with silicate features in moderate absorption 
(top panel, red line) in the RSA
     sample. The  shaded regions represent
     the $1\sigma$ dispersion.  The  three types clearly show
     nuclear $11.3\,\mu$m PAH emission. 
For comparison the grey line for each
     type is the mean circumnuclear {\it Spitzer}/IRS SL
     spectra.}\label{fig:average_MIRspectra}     
\end{figure}

Fig.~\ref{fig:average_MIRspectra} shows the resulting 
mean mid-IR spectra of RSA Seyfert 1 and Seyfert 2 nuclei, and Seyfert nuclei 
with silicate features in moderate absorption  compared to
their corresponding circumnuclear {\it Spitzer}/IRS SL spectra.  The 
nuclear spectra of Seyfert 1s and Seyfert 2s show 
similar mid-IR spectral indices (that is, ratio between 8 and
$12\,\mu$m), consistent with the results of \cite{Hoenig2010}
comparing individual galaxies. However, the  nuclear Seyfert
1 spectrum shows a flat or slightly in emission silicate feature
whereas the  Seyfert 2  shows the $10\,\mu$m feature slightly in absorption
\citep[see also][]{Hoenig2010}. 

The $11.3\,\mu$m PAH feature is detected in the average nuclear spectrum
of the three types.
Table~\ref{table:PAHfeature_composites} lists the
mean and standard deviation of the EW of the $11.3\,\mu$m PAH feature
for the average nuclear and IRS spectra for each type. We estimated the
uncertainties of the EW in the 
combined spectra using bootstrap resampling. To do so, first we selected at
random, and with replacement, a subset of the individual spectra used
to build the composite. The size of the resample was equal to the size
of the original set of spectra. We then obtained a composite spectrum for
this subset and measured the EW of the  feature. We repeated
this routine 100 times to get the bootstrap distribution of the EW. 

As expected, the nuclear
EWs of all three types are smaller than those measured in the
circumnuclear regions from 
the {\it Spitzer}/IRS spectra.  This is due to the higher contribution
of the mid-IR AGN continuum in the 
nuclear regions.  The measured EW of the $11.3\,\mu$m PAH feature of
Seyfert 1 and Seyfert 2 nuclei are similar, within the uncertainties. The
average spectrum of the nuclei with the silicate  
feature in moderate absorption  shows a slightly higher EW with a larger dispersion, but
again consistent with those 
of Seyfert 1 and 2 nuclei.  Since most of the nuclei with relatively deep
silicate features are  in highly
inclined systems there might be some contamination from
star formation in the line of 
sight of the host galaxy. Additionally, in these 
systems the silicate feature is likely to absorb 
the compact warm nuclear emission more than the diffuse PAH emission
which arises from a  much larger region.  This suppresses the $11\,\mu$m
continuum from the compact nuclear emission more than the $11.3\,\mu$m
PAH emission and an increased EW is expected in these cases. 

From this section and Section~6.1 we conclude that the decreasing EW of the
$11.3\,\mu$m PAH features of local AGN going from circumnuclear scales
(i.e. a few hundred pc probed by {\it 
  Spitzer}/IRS spectra) to nuclear 
scales of less than  100\,pc (i.e. ground-based data
obtained with $8-10\,$m class telescopes) is due to increased dilution from the
AGN continuum \citep[see also][]{RamosAlmeida2014}.  

\subsection{PAH survival in the vicinity of AGN?}\label{sec:PAHdestruction} 
The large molecules responsible for the long wavelength PAH emission
are expected 
to be less susceptible to destruction by 
X-ray photons produced by the AGN than those producing the 7.7 and
$8.6\,\mu$m PAH emission \citep{Smith2007}. Therefore, the $11.3\,\mu$m PAH 
feature is likely to be more resistant to the AGN radiation
field \citep[see also][]{DiamondStanic2010}. In this section we
evaluate whether there is a dependence of the detection of this feature
with the AGN luminosity and the projected distance of the star forming
regions from the AGN. 

 \begin{figure}

   \resizebox{0.97\hsize}{!}{\rotatebox[]{-90}{\includegraphics{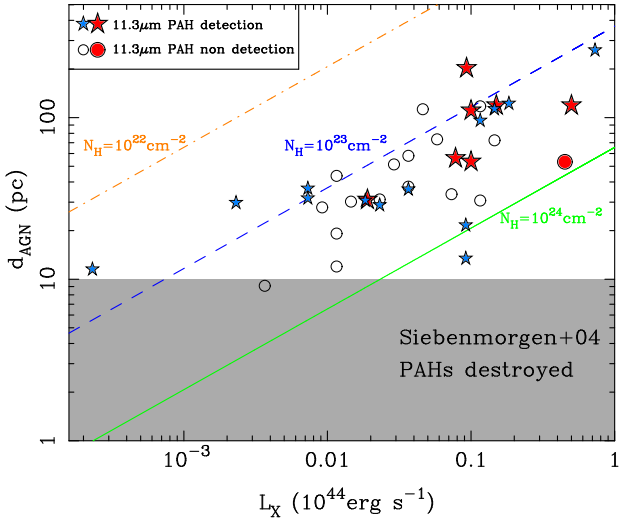}}}
\vspace{-0.5cm}

   \caption{Diagram showing regions of PAH survival as a function
     of the distance from the AGN and the AGN
     X-ray luminosity according to the modelling of Voit (1992) and 
Miles et al. (1994). The lines of constant total column densities (lines
from top to bottom 
     $N_{\rm H}{\rm (tot)}= 10^{22}, \, 10^{23}, \, 10^{24}\,$cm$^{-2}$) of the
     intervening material define the regions above which PAHs would be
     protected and not destroyed by the AGN radiation field. 
The shaded area shows the region where PAHs would not survive
($d_{\rm AGN}\le 10\,$pc) according to the modelling of 
Siebenmorgen et al. (2004). Galaxies with a nuclear
     detection of    the $11.3\,\mu$m PAH feature are shown as
     small and large star-like symbols for
   RSA galaxies from Esquej et al. (2014) and  this
   work, respectively. Circles are galaxies with non detections of the
   nuclear PAH feature. For the galaxies, the plotted X-ray
   luminosities are intrinsic 
   (corrected for absorption)  $2-10\,$keV
   luminosities and $d_{\rm AGN}$ are projected
   distances. }\label{fig:PAHdestruction}      
\end{figure}

According to the modeling of \cite{Voit1992}, to detect
PAH emission near AGN the rate of evaporation of the PAH carriers due to the
AGN radiation field has to be
lower than the rate at which carbon gets reaccreted onto the
PAHs. \cite{Voit1992} quantified the time scale needed for X-ray
reabsorption by a 50 carbon molecule
in terms of the distance of the PAH molecules from the
AGN $d_{\rm AGN}$, the AGN X-ray luminosity $L_X$, and the total hydrogen
column density of the 
intervening material $N_{\rm H}{\rm (tot)}$:  

\begin{equation}
\tau \sim 700 {\rm yr}\,\left(\frac{N_{\rm H}{\rm (tot)}}{10^{22}{\rm
      cm}^{-2}}\right)^{1.5}\, 
\left(\frac{d_{\rm AGN}}{{\rm kpc}}\right)^2\,\left(\frac{L_{\rm X}}{10^{44} {\rm erg\,s}^{-1}}\right)^{-1}.
\end{equation}

Fig.~\ref{fig:PAHdestruction} summarizes graphically the
predictions for PAH
survival, assuming a time scale for 
reaccretion of carbon onto a fractured PAH molecule of 3000\,yr
\citep[see][]{Voit1992,Miles1994}.  
We show lines of constant total hydrogen column densities as a function
of the AGN X-ray luminosity and distance from the AGN.  Below these
lines, PAH molecules might not survive the AGN radiation field. For a given
column density, the more luminous the AGN is, the further away the PAH
molecules ought to be in order to survive. Also, at a given AGN
luminosity, higher column densities of the intervening material
allow for smaller distances to the AGN of the star forming regions where the PAH
molecules could survive.
We also show in this figure for comparison the predictions from
\cite{Siebenmorgen2004} where the only dependence is the 
distance from the AGN, that is, PAH molecules would be destroyed at
distances of less than approximately 10\,pc from the AGN but survive
at distances of more than 100\,pc.

In Fig.~\ref{fig:PAHdestruction} we also plotted  the RSA Seyferts
from \cite{Esquej2014} and the galaxies from this work. For the  closest
galaxies in \cite{Esquej2014} sample  (Circinus, NGC~4945 and
NGC~5128) the distances from the AGN correspond to those probed by the {\it
  Spitzer}/IRS SL slit (see below). As can be seen from this figure, the majority of the
galaxies with nuclear $11.3\,\mu$m PAH detections 
cluster around the $N_{\rm H}{\rm (tot)}=10^{23}\,{\rm cm}^{-2}$  line and a
few just around the $N_{\rm H}{\rm (tot)}=10^{24}\,{\rm cm}^{-2}$
line.  

The derived total column densities from fitting the nuclear IR
emission of Seyfert nuclei with clumpy torus models 
are in this range \citep{RamosAlmeida2009,RamosAlmeida2011,Lira2013}.
Independently,
\cite{Hicks2009} derived typical column densities of 
$N_{\rm H} \sim 5 \times 10^{23}\,{\rm cm}^{-2}$ in the nuclear gas
disks (radii of 30\,pc) of Seyfert galaxies. Therefore, material 
in the dusty torus and/or gas disk has in principle sufficient column
density  to shield the PAH molecules from the AGN radiation field 
\citep[see also][]{Sales2013,Esquej2014}. Moreover, \cite{Hicks2009} proposed
that the molecular gas is mixed with the stellar population and is
likely to be associated with the outer regions of the obscuring
material (i.e. the torus).  This
scenario fits nicely with the physical scales where nuclear
$11.3\,\mu$m PAH feature emission is detected in this work (inner few
tens of pc)  and with  the  relatively compact 
sizes infered for the AGN tori
\citep{RamosAlmeida2009,RamosAlmeida2011,Tristram2009,AlonsoHerrero2011,Kishimoto2011,Burtscher2013}.

We can also see from Fig.~\ref{fig:PAHdestruction} that those nuclei where the
$11.3\,\mu$m PAH feature is not detected do  not lie in a particular
region of the diagram. Rather, they are mostly located between the lines
  of constant $N_{\rm H} = 10^{23}\,{\rm cm}^{-2}$ and $N_{\rm H} =
  10^{24}\,{\rm cm}^{-2}$, as those with nuclear detections of the
  feature. The question for the non-detections is 
whether they just do 
not have nuclear star formation or whether the PAH carriers have been
destroyed.  
For Circinus and NGC~5128, which are the nearest galaxies in
\cite{Esquej2014}, the nuclear PAH emission is not 
clearly detected at distances of less 10\,pc  from ground-based
spectroscopy \citep[see][respectively]{Roche2006, GonzalezMartin2013}.
The AGN of these two 
galaxies are  embedded, as implied by the deep nuclear $9.7\,\mu$m 
silicate features. In these cases it is not clear if the PAH
molecules are destroyed or not. For Circinus
\cite{Roche2006} concluded that  the $11.3\,\mu$m PAH emission
arises predominantly outside the regions affected by the large
columns producing the strong nuclear absorption and further from the
nucleus. For NGC~5128 \cite{TacconiGarman2013} detected nuclear
$3.3\,\mu$m PAH emission, and therefore it is possible that the
nuclear $11.3\,\mu$m PAH emission is embedded in the deep silicate feature. 

One interesting case of non detection of the nuclear $11.3\,\mu$m
  PAH feature 
is NGC~7213 \citep[see][]{Esquej2014, RuschelDutra2014}. The
nuclear spectrum of this Seyfert 1 galaxy shows the silicate feature
in emission. \cite{RuschelDutra2014} fitted the mid-IR spectra
with the \cite{Nenkova2008} models and derived a very low hydrodgen column
density ($N_{\rm H}\sim 10^{18}\,{\rm cm}^{-2}$) compared with the typical
values found in other Seyfert 1 and 
Seyfert 2 nuclei of $N_{\rm H} \sim 10^{23}-10^{24}\,{\rm
  cm}^{-2}$ \citep{RamosAlmeida2009, RamosAlmeida2011,
  RamosAlmeida2014, Sales2013, Sales2014,
  Lira2013, RuschelDutra2014}. 
For the AGN luminosity
of NGC~7213 and the distance to the AGN 
probed by the  high angular resolution mid-IR  
spectroscopy ($\log L(2-10{\rm keV})= 42.1\,{\rm
  erg \,s}^{-1}$ and
$d_{\rm AGN}=20\,$pc, see Esquej et 
al. 2014 and Ruschel-Dutra et al. 2014) the required column density to
protect the PAH molecules 
would be a few times $10^{23}\,{\rm cm}^{-2}$ (see
Fig.~\ref{fig:PAHdestruction}).  
The nuclear region of
NGC~7213 may represent a case where there is not sufficient protecting 
material in the dusty torus. We note  that this galaxy has a circumnuclear
ring of star formation  and some extended H$\alpha$ 
emission (a few arcseconds) around the nucleus  \citep[see
e.g.,][]{StorchiBergmann1996}. However, in the nuclear region of
NGC~7213 it is not clear if the H$\alpha$ emission line is
excited by nuclear  star formation activity,  by the AGN itself or both.

Finally,  some galaxies with a clear detection of the 
$11.3\,\mu$m PAH feature in the nuclear mid-IR spectrum
\citep[e.g. NGC~3783 and NGC~7469, see][]{Hoenig2010} do not show it in
 well-resolved interferometric observations probing much smaller physical scales
\citep[see][and references therein]{Burtscher2013}. The
second generation instruments planned for the VLTI will provide more
sensitive data and the possibility of image reconstruction. This will  be
useful to understand the behaviour of the nuclear $11.3\,\mu$m PAH
feature of AGN on smaller physical scales than those probed by mid-IR
spectroscopy on 8-10\,m class telescopes.

\section{Conclusions}
We  presented 
GTC/CanariCam Si-2 ($\lambda_{\rm c} = 8.7\,\mu$m) imaging and
$7.5-13\,\mu$m spectroscopy of six local 
systems (eight galaxies) known to host both an AGN and nuclear
(regions of less than 500\,pc) star formation activity. The main goal
of this work was to 
explore the behaviour of the  $11.3\,\mu$m PAH feature in the close
vicinity of an AGN. The high angular resolution ($0.24-0.40\,$arcsec)
of the GTC/CanariCam observations  allowed us to probe projected
distances from the AGN  between 30 and 210\,pc, depending on the
galaxy. We also  included  in our analysis the nuclear mid-IR
spectra of 29 RSA 
galaxies presented by \cite{Esquej2014}.  We summarize our
main conclusions as follows. 

\begin{enumerate}
\item We detected nuclear (inner $\sim 60-420\,$pc)
and extended (a few hundred pc from the AGN)  $11.3\,\mu$m PAH feature 
  emission in all the systems observed with GTC/CanariCam. 
The only
  exception is the nuclear region (inner $0.5\,{\rm arcsec} \sim 100$\,pc) of
  NGC~3690. However, the 
  upper limit of the EW of the feature is compatible with expectations from the
  comparison of its nuclear star formation rate with that of its companion
  galaxy IC~694.

\item We measured 
GTC/CanariCam nuclear  EW of the $11.3\,\mu$m PAH feature ranging from
values indicating   
AGN-dominated nuclear mid-IR emission (NGC~3690, NGC~2273, and Mrk~1073) to
intermediate cases (IC~694) to 
starburst-dominated nuclear mid-IR emission (both nuclei of NGC~6240,
Mrk~1066, and IRAS~17208$-$0014).  

\item The GTC/CanariCam spatially-resolved
spectroscopy shows that 
the $11.3\,\mu$m PAH feature emission peaks in the galaxy nuclei  with
the EW of the feature increasing further away  from the 
AGN. At projected distances from the AGN of a few 
hundred pc the EW of the feature reaches values consistent with
those of star-forming galaxies.

\item The nuclear EW of the $11.3\,\mu$m PAH feature are always lower 
than those measured from {\it Spitzer}/IRS spectra probing circumnuclear
regions, which are typically factors of 7 larger in size than the nuclear
regions. We explained the reduced nuclear EW of the $11.3\,\mu$m PAH
feature as due to an increased AGN continuum contribution in the
nuclear regions rather  than
destruction of the PAH carriers by the AGN radiation.

\item We investigated the survival of the PAH molecules responsible
  for the $11.3\,\mu$m PAH feature. We found no evidence of
  complete destruction of the  molecules responsible for this
   PAH feature in the nuclear 
regions of AGN at least to distances as close as 10\,pc from the
nucleus. 

\end{enumerate}

For Seyfert-like AGN luminosities and distances to the
nucleus probed by high angular resolution ($0.2-0.4\,$arcsec)  mid-IR
measurements the implied minimum column densities
of the intervening material needed to protect the molecules 
responsible for the $11.3\,\mu$m PAH feature are $N_{\rm H} = 10^{23} \,{\rm
  cm}^{-2}$ \citep[see also][]{Esquej2014}. Current modelling of the
nuclear IR emission of 
Seyfert galaxies with clumpy torus models indicates that the compact torus is
able to provide such column densities, even for most type 1
AGN \citep{RamosAlmeida2009, RamosAlmeida2011,
  AlonsoHerrero2011}. Likewise, column densities  
$N_{\rm H} \sim 5 \times 10^{23}\,{\rm cm}^{-2}$ are derived for
the nuclear gas disks of 
nearby AGN \citep{Hicks2009}. 

We conclude that the
$11.3\,\mu$m PAH feature can be used to probe 
nuclear star formation activity for Seyfert-like AGN luminosities and at least to
distances to the AGN of 10\,pc, assuming that the AGN contribution to the 
excitation of the $11.3\,\mu$m PAH feature is small or negligible.

\section*{Acknowledgments}

We thank Ric Davies for sharing with us  the VLT/SINFONI image of
NGC~6240. We are extremely grateful to the GTC staff for their
constant and enthusiastic support.  We thank the referee for
  comments that helped improve this work.

A.A.-H. and A.H.-C. are partly funded by the Universidad de Cantabria
through the Augusto G. Linares programme. A.A.-H. and
A.H.-C. acknowledge financial support from the Spanish Plan
Nacional grant AYA2012-31447,  A.A.-H. and P.E. from grant 
AYA2009-05705-E, C.R.A. from grant AYA2010-21887-C04.4 (Estallidos),
P.E. from grant AYA2012-31277, and L.C. from grant AYA2012-32295.
 C.R.A. acknowledges
financial support from the Marie Curie Intra European Fellowship
within the 7th European Community Framework Programme
(PIEF-GA-2012-327934) and 
S.F.H.  from the Marie Curie International
Incoming Fellowship within the 7th European Community Framework
Programme (PIIF-GA-2013-623804). The Dark Cosmology Centre is funded
by the DNRF. IA is partially funded by CONACyT grant
  SEP-CB-2011-01-167291. 
N.A.L. and R.E.M. are supported by the Gemini Observatory, which is
operated by the Association of Universities
for Research in Astronomy, Inc., on behalf of the international Gemini
partnership of
Argentina, Australia, Brazil, Canada, Chile, and the United States of
America. C.P. acknowledges support from UTSA to help enable this research.

Based on observations made with the GTC,
installed in the Spanish Observatorio del Roque de los Muchachos of the
Instituto de Astrof\'{\i}sica de Canarias, in the island of La Palma.
Based party on observations obtained with
the {\it Spitzer Space Observatory}, which is operated by JPL,
Caltech, under NASA contract 1407.
This research has made use of the NASA/IPAC Extragalactic Database
(NED) which is operated by JPL, Caltech, 
under contract with the National Aeronautics
and Space Administration. The Cornell Atlas of Spitzer/IRS
  Sources (CASSIS) is a product of the  
Infrared Science Center at Cornell University, supported by NASA and JPL.

\label{lastpage}


\begin{thebibliography}{99}
\bibitem[\protect\citeauthoryear{Aiken \&
  Roche}{1985}]{Aitken1985}Aitken D. K., Roche P. F., 1985, MNRAS, 213,  
777 
\bibitem[\protect\citeauthoryear{Alexander \&
  Hickox}{2012}]{Alexander2012}Alexander D. M.,  Hickox R. C., 
	2012, NewAR, 56, 93
\bibitem[\protect\citeauthoryear{Alonso-Herrero
  et al.}{2013a}]{AlonsoHerrero2013}Alonso-Herrero A., et al., 2013a,
ApJ, 779, L14
\bibitem[\protect\citeauthoryear{Alonso-Herrero
  et al.}{2013b}]{AlonsoHerrero2013LIRGs}Alonso-Herrero A., et al., 2013b,
ApJ, 765, 78
\bibitem[\protect\citeauthoryear{Alonso-Herrero
  et al.}{2012}]{AlonsoHerrero2012}Alonso-Herrero A.,
Pereira-Santaella M., Rieke G. H., Rigopoulou D., 2012, ApJ, 744, 2
\bibitem[\protect\citeauthoryear{Alonso-Herrero
  et al.}{2011}]{AlonsoHerrero2011}Alonso-Herrero A.,
et al., 2011, ApJ, 736, 82
\bibitem[\protect\citeauthoryear{Alonso-Herrero
  et al.}{2009}]{AlonsoHerrero2009}Alonso-Herrero A., Rieke G. H.,
Colina L., et al. 2009, ApJ, 697, 660
\bibitem[\protect\citeauthoryear{Alonso-Herrero
  et al.}{2006}]{AlonsoHerrero2006}Alonso-Herrero A., Colina L.,
Packham C., D\'{\i}az Santos T., Rieke G. H., Radomski J. T., Telesco
C., 2006, ApJ, 625, L83
\bibitem[\protect\citeauthoryear{Alonso-Herrero
  et al.}{2000}]{AlonsoHerrero2000}Alonso-Herrero A., Rieke G. H.,
Rieke M. J., Scoville N. Z., 2000, ApJ, 532, 845
\bibitem[\protect\citeauthoryear{Armus 
  et al.}{2006}]{Armus2006}Armus L., et al., 2006, ApJ, 640, 204
\bibitem[\protect\citeauthoryear{Arribas \&
  Colina}{2003}]{Arribas2003}Arribas S., Colina L., 2003, ApJ, 591, 791
\bibitem[\protect\citeauthoryear{Asmus
  et al.}{2014}]{Asmus2014}Asmus D., H\"onig S. F., Gandhi P., Smette
A., Duschl W. J., 2014, MNRAS, 439, 1648
\bibitem[\protect\citeauthoryear{Awaki 
et al.}{2009}]{Awaki2009}Awaki H., Terashima Y., Higaki Y., Fukazawa
Y., 2009, PASJ, 61, S317
\bibitem[\protect\citeauthoryear{Barbosa et
  al.}{2006}]{Barbosa2006}Barbosa F. K. B., Storchi-Bergmann T., Cid
Fernandes R., Winge C., Schmitt H., 2006, MNRAS, 371, 170
\bibitem[\protect\citeauthoryear{Beswick et
  al.}{2001}]{Beswick2001}Beswick R. J., Pedlar A., Mundell C. G.,
Gallimore J. F., 2001, MNRAS, 325, 151
\bibitem[\protect\citeauthoryear{Brandl et
  al.}{2006}]{Brandl2006}Brandl B. R., et al., 2006, ApJ, 653, 1129
\bibitem[\protect\citeauthoryear{Burtscher et
  al.}{2013}]{Burtscher2013}Burtscher L., et al., 2013, A\&A, 558, A149
\bibitem[\protect\citeauthoryear{Cid Fernandes et
  al.}{2001}]{CidFernandes2001}Cid Fernandes R., Heckman T., Schmitt
H., Gonz\'alez Delgado R. M., Storchi-Bergmann T., 2001, ApJ, 558, 81
\bibitem[\protect\citeauthoryear{Clavel et
  al.}{2000}]{Clavel2000}Clavel J., et al., 2000, A\&A, 357, 839
\bibitem[\protect\citeauthoryear{Davies et
  al.}{2007}]{Davies2007}Davies R. I., M\"uller S\'anchez F., Genzel
R., et al., 2007, ApJ, 671, 1388 
\bibitem[\protect\citeauthoryear{Della Ceca et
  al.}{2002}]{DellaCeca2002}Della Ceca R., et al., 2002, ApJ, 581, L9
\bibitem[\protect\citeauthoryear{Deo et
  al.}{2009}]{Deo2009}Deo R. P., Richards G. T., Crenshaw D. M.,
Kraemer S. B., 2009, ApJ, 705, 14
\bibitem[\protect\citeauthoryear{Diamond-Stanic \&
Rieke}{2012}]{DiamondStanic2012}Diamond-Stanic A. M., Rieke G. H.,
2012, ApJ, 746, 168
\bibitem[\protect\citeauthoryear{Diamond-Stanic \&
Rieke}{2010}]{DiamondStanic2010}Diamond-Stanic A. M.,  Rieke G. H.,
2010, ApJ, 724, 140
\bibitem[\protect\citeauthoryear{D\'{\i}az-Santos et al.
}{2010}]{DiazSantos2010}D\'{\i}az-Santos T., Alonso-Herrero A., Colina
L., Packham C., Levenson N. A., Pereira-Santaella M., Roche P. F.,
Telesco C., 2010, ApJ, 711, 328 
\bibitem[\protect\citeauthoryear{D\'{\i}az-Santos et al.
}{2008}]{DiazSantos2008}D\'{\i}az-Santos T., Alonso-Herrero A., Colina
L., Packham C., Radomski J., Telesco C., 2008, ApJ, 684, 211
\bibitem[\protect\citeauthoryear{Egami et
  al.}{2006}]{Egami2006}Egami E., Neugebauer G., Soifer B. T.,
Matthews K., Becklin E. E., Ressler M. E., 2006, AJ, 131, 1253
\bibitem[\protect\citeauthoryear{Engel et
  al.}{2010}]{Engel2010}Engel H., et al., 2010, A\&A, 524, A56
\bibitem[\protect\citeauthoryear{Esquej et
  al.}{2014}]{Esquej2014}Esquej P., Alonso-Herrero 
  A., Gonz\'alez-Mart\'{\i}n O., H\"onig S., Hern\'an-Caballero
  A., et al., 2014, ApJ, 780, 86
\bibitem[\protect\citeauthoryear{Ferruit et
  al.}{2000}]{Ferruit2000}Ferruit P., Wilson A. S., Mulchaey J., 2000,
ApJS, 128, 139
\bibitem[\protect\citeauthoryear{Genzel et
  al.}{1998}]{Genzel1998}Genzel R., et al. 1998, ApJ, 498, 579  
\bibitem[\protect\citeauthoryear{Gonz\'alez Delgado et
  al.}{2001}]{GonzalezDelgado2001}Gonz\'alez Delgado R. M., Heckman, T., 
Leitherer C., 2001, ApJ, 546, 845 
\bibitem[\protect\citeauthoryear{Gonz\'alez-Mart\'{\i}n et
  al.}{2013}]{GonzalezMartin2013}Gonz\'alez-Mart\'{\i}n O.,
  Rodr\'{\i}guez Espinosa J. M., D\'{\i}az-Santos T., et al., 2013,
  A\&A, 553, A35
\bibitem[\protect\citeauthoryear{Gonz\'alez-Mart\'{\i}n et
  al.}{2009}]{GonzalezMartin2009}Gonz\'alez-Mart\'{\i}n O., Masegosa
J., M\'arquez I., Guainazzi M., 2009, ApJ, 704, 1570
\bibitem[\protect\citeauthoryear{Goulding et
  al.}{2012}]{Goulding2012}Goulding A. D., et al., 2012, ApJ, 755, 5
\bibitem[\protect\citeauthoryear{Guainazzi et
  al.}{2005}]{Guainazzi2005}Guainazzi M., Matt G., Perola G. C., 2005,
A\&A, 444, 119
\bibitem[\protect\citeauthoryear{Helou et
    al.}{2004}]{Helou2004}Helou G., et al., 2004, ApJS, 154, 253
\bibitem[\protect\citeauthoryear{Hern\'an-Caballero \&
    Hatziminaoglou}{2011}]{HernanCaballero2011}Hern\'an-Caballero A., 
  Hatziminaoglou E., 2011, MNRAS, 414, 500 \
\bibitem[\protect\citeauthoryear{Hicks et 
    al.}{2009}]{Hicks2009}Hicks E. K. S., Davies R. I., Malkan M. A.,
  Genzel R., Tacconi L. J., M\"uller S\'anchez F., Sternberg A., 2009,
  ApJ, 696, 448 
\bibitem[\protect\citeauthoryear{H\"onig et
    al.}{2014}]{Hoenig2014}H\"onig S., Gandhi P., Asmus D., Mushotzky
  R. F., Antonucci R., Ueda Y., Ichikawa K., 2014, MNRAS, 438, 647
\bibitem[\protect\citeauthoryear{H\"onig et
    al.}{2010}]{Hoenig2010}H\"onig S., et al., 2010, A\&A, 515, 23
\bibitem[\protect\citeauthoryear{Hopkins \&
    Quataert}{2010}]{Hopkins2010}Hopkins P. F., Quataert E., 2010,
  MNRAS, 407, 1529 
\bibitem[\protect\citeauthoryear{Hopkins}{2012}]{Hopkins2012}Hopkins
  P. F., 2012, MNRAS, 420, L8
\bibitem[\protect\citeauthoryear{Houck et al.}{2004}]{Houck2004}Houck
  J. R., Roellig T. L., van Cleve  J., et al., 2004, ApJS, 154, 18 
\bibitem[\protect\citeauthoryear{Imanishi \&
Dudley}{2000}]{Imanishi2000}Imanishi M., Dudley C. C., 2000, ApJ, 545, 701
\bibitem[\protect\citeauthoryear{Kawakatu \&
Wada}{2008}]{Kawakatu2008}Kawakatu N., Wada K., 2008, ApJ, 681, 73
\bibitem[\protect\citeauthoryear{Kishimoto et
    al.}{2011}]{Kishimoto2011}Kishimoto M., H\"onig S. F., Antonucci
  R., et al., 2011, A\&A, 536, 78 
\bibitem[\protect\citeauthoryear{Komossa et
    al.}{2003}]{Komossa2003}Komossa S., Burwitz V., Hasinger G., et
  al., 2003, ApJ, 582, L15 
\bibitem[\protect\citeauthoryear{Laurent et
    al.}{2000}]{Laurent2000}Laurent O., et al., 2000, A\&A, 359, 887 
\bibitem[\protect\citeauthoryear{Lebouteiller et
    al.}{2011}]{Lebouteiller2011}Lebouteiller V., Barry D. J., Spoon
  H. W. W., et al., 2011, ApJS, 196, 8 
\bibitem[\protect\citeauthoryear{Lira et
    al.}{2013}]{Lira2013}Lira P., Videla L., Wu Y., Alonso-Herrero A.,
  Alexander D., Ward M., 2013, ApJ, 764, 59
\bibitem[\protect\citeauthoryear{Lutz et
    al.}{1998}]{Lutz1998}Lutz D., Spoon H. W. W., Rigopoulou D.,
  Moorwood A. F. M., Genzel D., 1998, ApJ, 505, L103
\bibitem[\protect\citeauthoryear{Maiolino et
    al.}{1995}]{Maiolino1995}Maiolino R., Ruiz M., Rieke G. H., Keller
  L. D., 1995, ApJ, 446, 561
\bibitem[\protect\citeauthoryear{Maiolino \&
   Rieke}{1995}]{MaiolinoRieke1995}Maiolino R., Rieke G. H., 1995,
 ApJ, 454, 95
\bibitem[\protect\citeauthoryear{Marinucci et
    al.}{2012}]{Marinucci2012}Marinucci A., Bianchi S., Nicastro F.,
  Matt G.,  Goulding A. D., 2012, ApJ, 748, 130
\bibitem[\protect\citeauthoryear{Max et
    al.}{2007}]{Max2007}Max C. E., Canalizo G., de Vries W. H., 2007,
  Science, 316, 1877
\bibitem[\protect\citeauthoryear{Miles et
    al.}{1994}]{Miles1994}Miles J. W., Houck J. R.,  Hayward, T. L.,
  1994, ApJ, 425, L37 
\bibitem[\protect\citeauthoryear{Mori et
    al.}{2014}]{Mori2014}Mori T. I., Imanishi M., Alonso-Herrero A.,
  et al., 2014, PASJ, in press (arXiv:1406.7780) 
\bibitem[\protect\citeauthoryear{Nardini et
    al.}{2008}]{Nardini2008}Nardini E., Risaliti G., Salvati M., Sani
  E., Imanishi M., Marconi A., Maiolino R., 2008, MNRAS, 385, L130
\bibitem[\protect\citeauthoryear{Nenkova et
    al.}{2008}]{Nenkova2008}Nenkova M., Sirocky M. M., Nikutta R.,
  Ivezi\'c Z., Elitzur M.,  2008, ApJ, 685, 160
\bibitem[\protect\citeauthoryear{Packham et
    al.}{2005}]{Packham2005}Packham C., et al., 2005, ApJ, 618, L17
\bibitem[\protect\citeauthoryear{Peeters et
    al.}{2004}]{Peeters2004}Peeters E., Spoon H. W. W., Tielens
  A. G. G. M., 2004, ApJ, 613, 986
\bibitem[\protect\citeauthoryear{Piqueras-L\'opez et
    al.}{2012}]{PiquerasLopez2012}Piqueras-L\'opez J., Colina L.,
  Arribas S., Alonso-Herrero A., Begregal A. G., 2012, A\&A, 546, A64
\bibitem[\protect\citeauthoryear{Quillen et
    al.}{1999}]{Quillen1999}Quillen A. C., Alonso-Herrero A., Rieke
  M. J., Rieke G. H., Ruiz M., Kulkarni V., 1999, ApJ, 527, 696
\bibitem[\protect\citeauthoryear{Radomski et
    al.}{2003}]{Radomski2003}Radomski J. T., Pi\~na R. K., Packham C.,
  Telesco C. M., De Buizer J. M., Fisher R. S., Robinson A., 2003, ApJ, 587, 117
\bibitem[\protect\citeauthoryear{Raimann et
    al.}{2003}]{Raimann2003}Raimann D., Storchi-Bergmann T.,
  Gonz\'alez Delgado R. M., Cid Fernandes R., Heckman T., Leitherer
  C., Schmitt H., 2003, MNRAS, 339, 772
\bibitem[\protect\citeauthoryear{Ramos Almeida et
    al.}{2009a}]{RamosAlmeida2009b}Ramos Almeida C., P\'erez
  Garc\'{\i}a A. M., Acosta-Pulido J. A., 2009a, ApJ, 694, 1379
\bibitem[\protect\citeauthoryear{Ramos Almeida et
    al.}{2009b}]{RamosAlmeida2009}Ramos Almeida C., Levenson N. A.,
  Rodr\'{\i}guez Espinosa J. M., et al., 2009b, ApJ, 702, 1127
\bibitem[\protect\citeauthoryear{Ramos Almeida et
    al.}{2011}]{RamosAlmeida2011}Ramos Almeida C., Levenson N. A.,
  Alonso-Herrero A., et al., 2011, ApJ, 731, 92
\bibitem[\protect\citeauthoryear{Ramos Almeida et
    al.}{2014}]{RamosAlmeida2014}Ramos Almeida C., 
  Alonso-Herrero A., et al., 2014, MNRAS, submitted
\bibitem[\protect\citeauthoryear{Riffel \&
    Storchi-Bergmann}{2011}]{Riffel2011}Riffel R. A., 
  Storchi-Bergmann T., 2011, MNRAS, 411, 469
\bibitem[\protect\citeauthoryear{Riffel et
    al.}{2010}]{Riffel2010}Riffel R. A., Storchi-Bergmann T., Nagar
  N. M., 2010, MNRAS, 404, 166
\bibitem[\protect\citeauthoryear{Roche et
    al.}{2006}]{Roche2006}Roche P. F., et al., 2006, MNRAS, 367, 1689
\bibitem[\protect\citeauthoryear{Roche et
    al.}{1991}]{Roche1991}Roche P. F., Aitken D. K., Smith C. H., Ward
  M. J., 1991, MNRAS, 248, 606
\bibitem[\protect\citeauthoryear{Ruschel-Dutra et
    al.}{2014}]{RuschelDutra2014}Ruschel-Dutra D., Pastoriza M.,
  Riffel R., Sales D. A., Winge C. 2014, 
MNRAS, 438, 3434
\bibitem[\protect\citeauthoryear{Sales et
    al.}{2014}]{Sales2014}Sales D. A., Ruschel-Dutra D., 
Pastoriza M. G.,  Riffel R., Winge C., 2014, MNRAS, 441, 630
\bibitem[\protect\citeauthoryear{Sales et
    al.}{2013}]{Sales2013}Sales D. A., Pastoriza M. G.,  Riffel R.,
Winge C., 2013, MNRAS, 429, 2634
\bibitem[\protect\citeauthoryear{Sales et
    al.}{2010}]{Sales2010}Sales D. A., Pastoriza M. G.,  Riffel R.,
  2010, ApJ, 725, 605
\bibitem[\protect\citeauthoryear{Sanders et
    al.}{2003}]{Sanders2003}Sanders D. B.,
  Mazzarella J. M., Kim D.-C., Surace J. A.,  Soifer B. T.,
  2003, AJ, 126, 1607  
\bibitem[\protect\citeauthoryear{Sani et
    al.}{2012}]{Sani2012}Sani E., et al., 2012, MNRAS, 424, 1936
\bibitem[\protect\citeauthoryear{Shi et
    al.}{2007}]{Shi2007}Shi Y., Ogle P., Rieke G. H., et al., 2007,
  ApJ, 669, 841 
\bibitem[\protect\citeauthoryear{Siebenmorgen \&
 Kr\"ugel}{2007}]{Siebenmorgen2007}Siebenmorgen R., Kr\"ugel E., 2007,
  A\&A, 461, 445 
\bibitem[\protect\citeauthoryear{Siebenmorgen 
 et al.}{2004}]{Siebenmorgen2004}Siebenmorgen R., Kr\"ugel E., Spoon
  H. W. W., 2004, A\&A, 414, 123 
\bibitem[\protect\citeauthoryear{Smith et
    al.}{2007}]{Smith2007}Smith J. D., Draine B., Dale D., et al., 2007,
  ApJ, 656, 770 
\bibitem[\protect\citeauthoryear{Smith et
    al.}{1996}]{Smith1996}Smith D. A., Herter T., Haynes M. P.,
  Beichman C. A., Gautier T. N., 1996, ApJS, 104, 217
\bibitem[\protect\citeauthoryear{Soifer et
    al.}{2000}]{Soifer2000}Soifer B. T., et al., 2000, AJ, 119, 509
\bibitem[\protect\citeauthoryear{Spoon et
  al.}{2007}]{Spoon2007}Spoon H. W. W., Marshall J. A., Houck J. R.,
Elitzur M., Hao L., Armus L., Brandl B. R., Charmandaris V., 2007, ApJ, 654, 49
\bibitem[\protect\citeauthoryear{Storchi-Bergmann et al.
    }{1996}]{StorchiBergmann1996}Storchi-Bergmann T., Rodriguez-Ardila
    A., Schmitt H. R., Wilson 
A. S., Baldwin J. A. 1996, ApJ, 472, 83
\bibitem[\protect\citeauthoryear{Tacconi-Garman \&
    Sturm}{2013}]{TacconiGarman2013}Tacconi-Garman L. E., Sturm E.,
  2013, A\&A, 551, A139
\bibitem[\protect\citeauthoryear{Tacconi-Garman et
    al.}{2005}]{TacconiGarman2005}Tacconi-Garman L. E., Sturm E.,
  Lehnert M., Lutz D., Davies R. I., Moorwood A. F. M., 
  2005, A\&A, 432, 91
\bibitem[\protect\citeauthoryear{Telesco et
    al.}{2003}]{Telesco2003}Telesco C. M., Ciardi
 D., French J., et al., 2003, in SPIE Conference Series, Vol. 4841, 913
\bibitem[\protect\citeauthoryear{Tielens
    }{2013}]{Tielens2013}Tielens A. G. G. M., 2013, RvMP, 85, 1021
\bibitem[\protect\citeauthoryear{Tommasin et al.
    }{2010}]{Tommasin2010}Tommasin S., Spinoglio L., Malkan M. A.,
    Fazio G., 2010, ApJ, 709, 1257
\bibitem[\protect\citeauthoryear{Tristram et
    al.}{2009}]{Tristram2009}Tristram K. R. W., Raban D., Meisenheimer
  K., et al., 2009, A\&A, 502, 67 
\bibitem[\protect\citeauthoryear{Veilleux et
    al.}{1997}]{Veilleux1997}Veilleux S., Goodrich R. W., Hill G. J.,
  1997, ApJ, 477, 631
\bibitem[\protect\citeauthoryear{Voit}{1992}]{Voit1992}Voit
  G. M., 1992, MNRAS, 258, 841 
\bibitem[\protect\citeauthoryear{Vollmer 
et al.}{2008}]{Vollmer2008}Vollmer B., Beckert T., Davies R. I., 2008,
A\&A, 491, 441
\bibitem[\protect\citeauthoryear{Wada
\& Norman}{2002}]{Wada2002}Wada K.,  Norman C. A., 2002, ApJ, 566, L21 
\bibitem[\protect\citeauthoryear{Wild
et al.}{2010}]{Wild2010}Wild V., Heckman T., Charlot S., 2010, MNRAS,
405, 933
\bibitem[\protect\citeauthoryear{Wu
et al.}{2009}]{Wu2009}Wu Y., Charmandaris V., Huang J., Spinoglio L.,
Tommasin S., 2009, ApJ, 701, 658
\end{thebibliography}
\end{document}